\documentclass[ams,prd,superscriptaddress,preprintnumbers,eqsecnum,nofootinbib,notitlepage]{revtex4-2}
\usepackage{graphicx} % Required for inserting images
\usepackage{amsmath,amssymb,amsfonts}
\usepackage{xcolor}

\begin{document}

\preprint{YITP-23-144, IPMU23-0043}
\title{Observational bounds on extended minimal theories of massive gravity: New limits on the graviton mass}

\author{Antonio De Felice}
\email{antonio.defelice@yukawa.kyoto-u.ac.jp}
\affiliation{
Center for Gravitational Physics and Quantum Information, Yukawa Institute for Theoretical Physics, Kyoto University, 606-8502, Kyoto, Japan
}

\author{Suresh Kumar}
\email{suresh.math@igu.ac.in}
\affiliation{Department of Mathematics, Indira Gandhi University, Meerpur, Haryana 122502, India}

\author{Shinji Mukohyama}
\email{s.mukohyama@gmail.com}
\affiliation{
Center for Gravitational Physics and Quantum Information, Yukawa Institute for Theoretical Physics, Kyoto University, 606-8502, Kyoto, Japan
}
\affiliation{Kavli Institute for the Physics and Mathematics of the Universe (WPI), The University of Tokyo Institutes for Advanced Study, The University of Tokyo, Kashiwa, Chiba 277-8583, Japan}

\author{Rafael C.\ Nunes}
\email{rafadcnunes@gmail.com}
\affiliation{Instituto de F\'{i}sica, Universidade Federal do Rio Grande do Sul, 91501-970 Porto Alegre RS, Brazil}
\affiliation{Divis\~ao de Astrof\'isica, Instituto Nacional de Pesquisas Espaciais, Avenida dos Astronautas 1758, S\~ao Jos\'e dos Campos, 12227-010, SP, Brazil}

\date{\today}
%======================================%
%<<<<<<<<<<<<< ABSTRACT >>>>>>>>>>>>>>>%
%======================================%
\begin{abstract}
In this work, we derive for the first time observational constraints on the extended Minimal Theory of Massive Gravity (eMTMG) framework in light of Planck-CMB data, geometrical measurements from Baryon Acoustic Oscillation (BAO), Type Ia supernovae from the recent Pantheon+ samples, and also using the auto and cross-correlations cosmic shear measurements from KIDS-1000 survey. Given the great freedom of dynamics choice for the theory, we consider an observationally motivated subclass in which the background evolution of the Universe goes through a transition from a (positive or negative) value of the effective cosmological constant to another value. From the statistical point of view, we did not find evidence of such a transition, i.e. deviation from the standard $\Lambda$CDM behavior, and from the joint analysis using Planck + BAO + Pantheon+ data, we constrain the graviton mass to $< 6.6 \times 10^{-34}$ eV at 95\% CL.  We use KIDS-1000 survey data to constrain the evolution of the scalar perturbations of the model and its limits for the growth of structure predicted by the eMTMG scenario. In this case, we find small evidence at 95\% CL for a non-zero graviton mass. We interpret and discuss these results in light of the current tension on the $S_8$ parameter. We conclude that, within the subclass considered, the current data are only able to impose upper bounds on the eMTMG dynamics. Given its potentialities beyond the subclass, eMTMG can be classified as a good candidate for modified gravity, serving as a framework in which observational data can effectively constrain (or confirm) the graviton mass and deviations from the standard $\Lambda$CDM behavior. 
\end{abstract}

\maketitle

\section{Introduction}

The nature of the current stage of accelerated expansion of the Universe remains one of the main open problems in modern astronomy \cite{DE_obs_1, DE_obs_2}. The $\Lambda$-Cold Dark Matter scenario (the $\Lambda$CDM scenario) is the simplest hypothesis which explains the current late-time accelerated expansion of the Universe while being able to explain the majority of the latest observations both at astrophysical and cosmological scales. At the theoretical level, because of the cosmological constant problem \cite{Weinberg:1988cp, Padmanabhan:2002ji, Copeland_2006}, the community has been looking for alternatives to the $\Lambda$ term. One may consider extra degrees of freedom with a gravitational origin, i.e., arising from a gravitational modification that possesses general relativity (GR) as a particular limit. The modified gravity (MG) scenarios may allow for extensions of the $\Lambda$CDM model which exhibit the accelerated expansion of the Universe at late times, as well as explain various observations at the cosmological and astrophysical scales (see \cite{Ishak:2018his, CANTATA:2021ktz, Heisenberg_2019} for a recent review).

On the other hand, recently a few tensions and anomalies have turned out statistically significant while analyzing different data sets.  The most statistically significant disagreement is in the value of the Hubble constant, $H_0$, that is between the Planck Cosmic Microwave Background (CMB) estimate \cite{Planck:2018vyg}, assuming the standard $\Lambda$CDM model, and the direct local distance ladder measurements conducted by the SH0ES team \cite{Riess:2021jrx, Riess_2022,murakami2023leveraging}, reaching a significance of more than 5$\sigma$. Further, within the $\Lambda$CDM framework, the CMB measurements from Planck and ACT-DR6 \cite{Planck:2018vyg,qu2023atacama} provide values of $S_8 = \sigma_8 \sqrt{\Omega_m/0.3}$ in 2-3$\sigma$ statistical tension with the ones inferred from various weak lensing, galaxy clustering, and redshift-space distortion measurements \cite{DiValentino:2020vvd, Nunes_2021, DES:2021bvc, Asgari_2021}. Various other anomalies and tensions have been emerging within the $\Lambda$CDM framework in recent years \cite{Perivolaropoulos_2022}. Motivated by such discrepancies, unlikely to disappear completely by introducing multiple systematic errors, it has been widely discussed in the literature whether new physics beyond the standard cosmological model can solve these tensions, and theories beyond GR can serve as alternatives to explain these current tensions \cite{DiValentino:2021izs}.

One of the most interesting possibilities for modifications of gravity is to give a mass to the graviton (see \cite{dRGT_2} for a review of massive gravity theories). Motivated by the potentialities of this framework,  a new theory of massive gravity with only two propagating degrees of freedom, namely the minimal theory of massive gravity (MTMG), was presented in \cite{MTMG:origpap, MTMG:pheno}. The MTMG framework has been tested and explored in the astrophysical and cosmological contexts in \cite{deAraujo:2021cnd, DeFelice:2021trp, MTMG:isw, MTMG:BH}. The MTMG formulation with a quasidilaton scalar field has been developed in \cite{MQD_OP, MQD_Horndeski, MQD_pheno}, and several other classes of the minimally modified gravity have been proposed and tested \cite{DeFelice:2020ecp, MMTG_OP, MMTG_Carballo_Rubio, MMTG_Hamiltonian, MMTG_Planck, VCDM:solvingH0, Yao:2020tur}. Recently, the \texttt{extended minimal theory of massive gravity} (eMTMG) was introduced in \cite{DeFelice:2022mcd}. The eMTMG is constructed to have only two degrees of freedom in the gravity sector, and free of the ghosts and/or instabilities. At the cosmological level, the theory predicts that the effective gravitational constant is always finite (a feature that was missing in MTMG) but not equal to unity in general, to allow some non-trivial modifications of gravity, besides the massive tensorial modes, and imposes that the square of mass of the graviton is always non-negative.

In this work, we aim to constrain the eMTMG for the first time using the latest cosmological data under the following considerations. The recent developments in \cite{Akarsu_2021, Akarsu_2023,akarsu2023lambdarm}, where a transition of the Universe from (anti-)de Sitter\footnote{More in general, we will not set a prior the sign of the effective cosmological constant at early times. Therefore, the theory also allows a transition from two different de Sitter vacua.} vacuum to a de Sitter one is postulated to exist, are inspiring and promising from an observational point of view. In this work, we present a physically well-motivated theory framework that allows shifting between negative and positive cosmological constants' phases within the framework of eMTMG. Then, we derive some observational bounds on the massive gravity modes. We discuss whether our proposal is viable or not, and how to improve the model in light of recent cosmological tensions.

The eMTMG has the feature that the background evolution is not fixed a priori, but once it is chosen, the theory is completely predictive. This is analogous to fixing a potential for a quintessence model. However, the theory by construction only possesses two gravitational degrees of freedom, i.e., two massive gravitational waves. In this paper, we aim to find bounds on the mass of the tensor modes from existent cosmological data sets and explore the possibilities of the free background in the context of the $H_0$ and $S_8$ tensions. We find the data do constrain the value of the mass for the tensor modes. However, we also find that to solve the above-mentioned tensions a simple shift in the cosmological constant at low and high redshifts is not a sufficient modification from the $\Lambda$-CDM background. We need a search running to explore the set of functions that determine the background dynamics, but we leave it for future work.

This paper is structured as follows. In Section \ref{sec:model}, we review and introduce the eMTMG scenario, whereas treatment of the background and perturbations dynamics is discussed in Section \ref{sec:back} and \ref{sec:perturbations} respectively. In Section \ref{data}, we present the data sets and methodology used in this work. In section \ref{results}, we discuss the main results of our analysis. In section \ref{final}, we summarize the main findings of this study.

\section{The model}\label{sec:model}

In the following, we will consider eMTMG, introduced in \cite{DeFelice:2022mcd}, which is a theory having only two gravitational degrees of freedom as in General Relativity (GR), with the difference that the model breaks four-dimensional diffeomorphism invariance in the unitary gauge. To have only two gravitational degrees of freedom and not the five ones of the ghost-free massive gravity (dRGT) introduced in \cite{dRGT_1,dRGT_2}, we need to define the theory through four constraints that can keep the two tensor modes as the only propagating degrees of freedom on any background. This procedure was first implemented in \cite{MTMG:origpap} to remove the instabilities present in dRGT on a homogeneous and isotropic background \cite{DeFelice:2012mx, DeFelice:2013awa}. This last theory was called minimal, the minimal theory of massive gravity (MTMG).

Then what does eMTMG add to the above-mentioned MTMG model of massive gravity? Similarly to MTMG, it only possesses two local gravitational degrees of freedom. The construction algorithm for eMTMG is essentially the same as MTMG, as it makes use of four constraints to keep only the two propagating local physical degrees of freedom in the gravity sector, that correspond to two massive tensor modes. However, the construction follows also other goals and its phenomenology is quite different. 

In the phenomenology of MTMG, on homogeneous and isotropic backgrounds, the effective gravitational constant felt by the cosmological linear perturbation fields can be written as
\begin{equation}
\label{Geff}
    \frac{G_{\rm eff}}{G_N} = \frac1{1-\tfrac12{\tilde\theta} Y}-\frac1{({\tilde\theta}Y-2)^2}\,\frac{\rho_m}{M_{\rm P}^2H^2}\,{\tilde\theta}Y\,,
\end{equation}
where $\tilde\theta \equiv \mu^2/H_0^2$, $\mu^2$ is the squared mass of the tensor modes, and $Y\equiv H_0^2/H^2=3M_{\rm P}^2H_0^2/(\rho_X+\rho_m)$. It is clear that in the limit $\mu^2\to0$, then $\frac{G_{\rm eff}}{G_N}\to1$. The same limit is also reached for $Y\to0$, that is at early times, or whenever $\rho_m\gg\rho_X$. However, there will be times or even regions of space (e.g.\ in the cosmological voids) for which $\rho_m\simeq\rho_X$. In that case, the quantity $\tilde\theta Y$ could be of order unity, and as such the denominator of $\frac{G_{\rm eff}}{G_N}$ could vanish. In this case, the effective squared mass of the matter density fluctuations would diverge, and before that happens, the theory goes out of the validity of a well-defined EFT approach. Therefore a theory that, at the beginning, seems to be able to give a non-trivial phenomenology for the growth of structure makes the deviation from GR too drastic, by reaching points outside the EFT validity.

Therefore the model(s) of eMTMG have been introduced, exactly to prevent this phenomenon from happening, while keeping a non-trivial phenomenology in terms of structure formation, easily allowing, for instance, $0<\frac{G_{\rm eff}}{G_N}\leq1$, an ingredient which may turn out indispensable if future experiments will confirm and exacerbate the $S_8$ tension in the $\Lambda$CDM model.

The expression for the most general eMTMG models is not known so far, however, in \cite{DeFelice:2022mcd}, one of its simplest realizations was explicitly given. In this paper, we will focus on such a subset of the eMTMG model, which is described by five parameters more than $\Lambda$CDM together, as we will see, with a free function that, on homogeneous and isotropic manifolds, corresponds to a free function of time.

First of all, we introduce two functions
\begin{align}
F_1(\mathcal{A},\mathcal{B})&=
2c_4+\xi^2\,\mathcal{A}\,\Bigl(2\mathcal{B}-\tfrac{10}9\mathcal{A}^2\Bigr)\,,\label{eq:F1}\\
F_2(A,B,C)&=\zeta_{1}^{2} \left(2 A B -\frac{10}{9} A^{3}\right)+\zeta_{2}^{2} \left(2 A C -\frac{4}{9} A^{4}\right)+\zeta_{3}^{2} \left(2 B C -\frac{2 A^{5}}{15}\right)+\zeta_{4}^{2} \left(C^{2}-\frac{A^{6}}{45}\right),
    \label{eq:F2}
\end{align}
where $c_4$ plays the role of a bare cosmological constant, whereas $\xi$ and $\zeta_i$ ($i\in\{1,2,3,4\}$) are real numbers. To understand the variables on which $F_{1,2}$ depend, we need to introduce the concept of the three-dimensional fiducial metric, ${\tilde\gamma}_{ij}$, which, in the unitary gauge is written and chosen as ${\tilde\gamma}_{ij}={\tilde a}^2(t)\,\delta_{ij}$. On considering instead the physical metric, in particular, in its ADM setup, we can introduce its three-dimensional components as $\gamma_{ij}(t,\vec{x})$. Then out of these two three-dimensional metrics, of which only the latter has still unknown dynamics that should be determined employing the equations of motion of eMTMG, one can build up the following three-dimensional tensors $\mathcal{K}^i{}_j$ and $\mathfrak{K}^i{}_j$ as
\begin{align}
    \mathcal{K}^i{}_l\,\mathcal{K}^l{}_j&={\tilde\gamma}^{il}\gamma_{lj}\,,\\
    \mathfrak{K}^i{}_l\,\mathfrak{K}^l{}_j&={\gamma}^{il}{\tilde\gamma}_{lj}\,,\\ \mathcal{K}^i{}_l\,\mathfrak{K}^l{}_j&=\delta^i{}_j=\mathfrak{K}^i{}_l\,\mathcal{K}^l{}_j\,,
\end{align}
where $\tilde{\gamma}^{ij}$ and $\gamma^{ij}$ correspond to the inverse of $\tilde{\gamma}_{ij}$ and $\gamma_{ij}$ respectively. We are now in the position to define the following quantities 
\begin{alignat}{4}
    \mathcal{A}&=[\mathfrak{K}]\,,\qquad&\mathcal{B}&=[\mathfrak{K}^2]\,,\\
    A&=[\mathcal{K}]\,,& B&=[\mathcal{K}^2]\,,\qquad C&=[\mathcal{K}^3]\,,
\end{alignat}
where $[\mathcal{K}]\equiv \mathcal{K}^i{}_i$, $[\mathcal{K}^2]\equiv \mathcal{K}^i{}_l\,\mathcal{K}^l{}_i$, etc.

In the following, we will make use of the ADM splitting to write down the theory. In particular, we will split the full Lagrangian density into several contributions. The first one is the Einstein-Hilbert term, namely
\begin{equation}
    \mathcal{L}_1\equiv\mathcal{L}_{\rm EH}=\frac{M_{\rm P}^2}2\,N\sqrt\gamma\,(K_{ij}K^{ij}-K^2+R)\,,
\end{equation}
where $N$ is the lapse function, whereas $\gamma$ represents the determinant of $\gamma_{ij}$. We made also use of the three-dimensional extrinsic curvature tensor, $K_{ij}$, defined by
\begin{equation}
    K_{ij}=\frac1{2N}\,({\dot\gamma}_{ij}-\gamma_{ik}D_jN^k-\gamma_{jk}D_iN^k)\,,
\end{equation}
where $N^i$ is the three-dimensional shift-vector for the physical metric, and the operator $D_i$ represent the three-dimensional covariant derivative compatible with the metric $\gamma_{ij}$. Then we also have that $K=K_{ij}\gamma^{ij}$, $K^{ij}=\gamma^{il}\gamma^{jm}K_{lm}$, and $R$ is the three-dimensional Ricci scalar for the metric $\gamma_{ij}$.

At this point, we can add two contributions to the Lagrangian density as
\begin{equation}
    \mathcal{L}_2 = -\frac12\,m^2M_{\rm P}^2\sqrt\gamma\,N\,F_1
    -\frac12\,m^2M_{\rm P}^2\sqrt{\tilde\gamma}\,M\,F_2\,,\label{eq:L2}
\end{equation}
where the two functions $F_{1,2}$ have been already defined above in Eqs.\ \eqref{eq:F1} and \eqref{eq:F2}. In Eq.\ \eqref{eq:L2} we have also made use of the fiducial lapse $M=M(t)$, which needs to be thought, in the unitary gauge, as a given function of time.

We continue building the Lagrangian by introducing the following symmetric tensor
\begin{equation}
    \Theta^{ij}=-\frac{\sqrt{\tilde\gamma}}{\sqrt\gamma}\,[({\tilde\gamma}^{jk}\mathfrak{K}^i{}_k+{\tilde\gamma}^{ik}\mathfrak{K}^j{}_k)F_{2,A}
    +4{\tilde\gamma}^{ij}F_{2,B}
    +3({\tilde\gamma}^{jk}\mathcal{K}^i{}_k+{\tilde\gamma}^{ik}\mathcal{K}^j{}_k)F_{2,C}]\,,
\end{equation}
out of which we have another contribution for the Lagrangian
\begin{equation}
    \mathcal{L}_3=\frac{m^4M_{\rm P}^2}{64}\frac{M^2}{N^2}\,\lambda^2\,N\sqrt\gamma\,(2\Theta_{ij}\Theta^{ij}-\Theta^i{}_{i}\,\Theta^j{}_{j}),\label{eq:L3}
\end{equation}
where the indices of $\Theta^{ij}$ have been lowered by means of $\gamma_{ij}$. In Eq.\ \eqref{eq:L3} we have introduced the Lagrange multiplier $\lambda$, which is one of the fields that make the theory minimal, i.e.\ having only two degrees of freedom in the gravitational sector.

Next, we notice that out of the fiducial metric and lapse, one can introduce the following fiducial tensor
\begin{equation}
    {\bar\zeta}^i{}_j\equiv\frac1{2M}\,{\tilde\gamma}^{il}\dot{\tilde\gamma}_{lj} = \frac{\dot{\tilde a}}{M{\tilde a}}\,\delta^i{}_j\,.
    \label{eq:zeta_ij}
\end{equation}
We are now in the position to introduce the following building block
\begin{equation}
   \mathcal{C}_\zeta \equiv -\frac12\,m^2M_{\rm P}^2\,M\,{\bar\zeta}^j{}_i\bigl(
   F_{1,\mathcal{A}}\gamma^{ik}\mathcal{K}^l{}_k{\tilde\gamma}_{lj} 
   + 2F_{1,\mathcal{B}}\gamma^{ik}{\tilde\gamma}_{kj}\bigr)\,,
\end{equation}
out of which the following contribution is found
\begin{equation}
    \mathcal{L}_4=\lambda\sqrt\gamma\,\bigl(\mathcal{C}_\zeta-\tfrac14\,m^2M_{\rm P}^2MK_{ij}\Theta^{ij}\bigr)\,.\label{eq:L4}
\end{equation}
To make the theory minimal other three constraints need to be imposed. First of all, let us introduce the following tensor
\begin{equation}
    \mathcal{C}^i{}_j=\frac12\,m^2\,M_{\rm P}^2\,M\,\frac{\sqrt{\tilde\gamma}}{\sqrt\gamma}\left[\frac12\,F_{2,A}(\mathfrak{K}^i{}_l{\tilde\gamma}^{lk}+
    {\tilde\gamma}^{il}\mathfrak{K}^k{}_l)\gamma_{kj}
    +2F_{2,B}{\tilde\gamma}^{ik}\gamma_{kj}
    +\frac32\,F_{2,C}(\mathcal{K}^i{}_l{\tilde\gamma}^{lk}
    +{\tilde\gamma}^{il}\mathcal{K}^{k}{}_{l})\gamma_{kj}\right],
\end{equation}
out of which, we can introduce the last gravitational contribution to the Lagrangian density
\begin{equation}
    \mathcal{L}_5=\sqrt{\gamma}\,\mathcal{C}^i{}_j\,D_i\lambda^j\,,\label{eq:L5}
\end{equation}
where we have introduced a three-dimensional Lagrange multiplier $\lambda^i$.

Finally, we can build up the total action for the theory
\begin{equation}
    S=\int d^4x \sum_{a=1}^5 \mathcal{L}_a + S_{\mathrm{mat}}\,,\label{eq:action}
\end{equation}
where the matter action is the same as in General Relativity.

\section{The background}\label{sec:back}

We will consider the behavior of the extended minimal theory of massive gravity (eMTMG) as introduced in Sec.\ \ref{sec:model} on a spatially flat homogeneous and isotropic background.

The fiducial sector already follows the high level of symmetries imposed by the background, namely $M=M(t)$ and $\tilde\gamma_{ij}={\tilde a}^2\,\delta_{ij}$. We assume that the physical variables defining the four-dimensional metric at the background level also follow the standard homogeneous and isotropic ansatz
\begin{align}
    N &= N(t)\,,\\
    N^i &= 0\,,\\
    \gamma_{ij} &= a(t)^2\,\delta_{ij}\,.
\end{align}
But to define eMTMG we also need to give a background profile for the three-dimensional Lagrange multiplier
\begin{equation}
    \lambda^i = 0\,,\qquad \lambda=\lambda(t)=0\,.
\end{equation}
If it is clear that the symmetry of the background imposes that the three-dimensional vector should have a zero background profile, it is possible (but non-trivial) to show that the background equations of motion lead to $\lambda(t)=0$ as the only meaningful profile for the field $\lambda$, see \cite{DeFelice:2022mcd}. In short, from a Lagrangian point of view, this profile is the only one that allows all the equations of motion to close without imposing extra constraints which would drastically reduce the allowed dynamics for the background.
For each matter field, labeled with the index ``$I$,'' we have instead, as usual, that
\begin{equation}
    \rho_I=\rho_I(t)\,,\quad P_I=P_I(t)\,,\quad {\dot\rho}_I +3 \dot{a} (\rho_I+P_I)/a = 0\,,
\end{equation}
which requires an equation of state, e.g.\ $P_I=P_I(\rho_I)$, to be solved.

In the following, it is convenient to introduce the quantity $X=X(t)$, which corresponds to the ratio of the fiducial scale factor to the one of the physical metric, that is $X\equiv {\tilde a}(t)/a(t)$. The dynamics of such a function are not determined a priori because we have the freedom to pick up any wanted dynamics for the fiducial scale factor $\tilde{a}(t)$. Nonetheless, from now on, we will impose that $X>0$. The choice of $X(t)$ determines not only the background dynamics but also influences the behavior of the perturbations. Therefore, even inside the same eMTMG set, different choices of $X(t)$ will correspond not only to different phenomenology but also to different fiducial metrics, i.e.\ different theories.

As was shown in \cite{DeFelice:2022mcd}, we can write down the Friedmann equation of motion for the background as 
\begin{align}
    3M_{\rm P}^2 H^2 &= \sum_I\rho_I + \rho_X\,,\label{eq:Frd1}\\
    \rho_X &\equiv m^2M_{\rm P}^2 (c_4-6\xi^2X^3)\,,
\end{align}
where $H\equiv \dot{a}/(aN)$, and the sum over $I$ in Eq.\ \eqref{eq:Frd1} runs over all the standard matter fields (including, for instance, the cold dark matter component).

Since we also have a fiducial lapse, it is also convenient to introduce another function, $r=r(t)$, which is defined as
\begin{equation}
    r \equiv \frac1X\,\frac{M}{N}\,,
\end{equation}
and which is required to be positive during the entire evolution of our universe, because $M$ and $N$ are lapse functions for the fiducial and the physical metric respectively. At this level, the choice of the function $M(t)$ is an additional ingredient of eMTMG, independent of the choice for $X(t)$.

However, for the eMTMG defined by the action Eq.\ \eqref{eq:action}, on the homogeneous and isotropic background, we have the following constraint holds:
\begin{equation}
    X'+X = \frac{5\zeta_1^2X^3 +10\zeta_2^2X^2 +10\zeta_3^2 X +6\zeta_4^2}{5\xi^2 X^4}\, r\,,\label{eq:constr}
\end{equation}
where a prime denotes differentiation with respect the e-fold variable $\mathcal{N}=\ln(a/a_{0})$.
This constraint corresponds to the equation of motion imposed by the field $\lambda$. This equation can be interpreted and used in two ways: 1) the dynamics of $X(t)$ are imposed by hand, as we will do in the following, and then the equation fixes the dynamics for $r(t)$ or 2) the dynamics of $r(t)$ is given, and Eq.\ \eqref{eq:constr} consists of an ODE for $X(t)$. In both cases, one needs to make sure that in the process both $X(t)$ and $r(t)$ remain positive during the evolution of our universe. Therefore not all positive $X(t)$ satisfy Eq.\ \eqref{eq:constr}, i.e.\ leading a positive $r(t)$. This implies that only a subset of all possible dynamics for $X(t)$ are consistent with eMTMG.

As we will see later on, the following background quantity determines the squared mass of the tensor modes:
\begin{equation}
    \mu^2 = 6m^2 r\,\bigl(\zeta_2^2 +2\zeta_3^2/X + \tfrac95\,\zeta_4^2/X^2\bigr)\,,
\end{equation}
which is a priori a non-negative quantity, as it can vanish for $\zeta_i=0$ ($i\in{2,3,4}$). The fact that eMTMG leads to $\mu^2\geq0$ for any background dynamics of $X(t)>0$ is exactly one of the phenomenological requirements imposed to define eMTMG. Out of this quantity, which is a function of time, we can define the following parameter $\theta_0\equiv \mu(t=0)/H_0$ which is, by construction, a non-negative real number.

%%%%%%%%%%%%%%
As we have already stated above, to completely define eMTMG and, hence, to be able to make any prediction, we need to specify the functional form of $X(t)$ (or $r(t)$, as stated above). 

This is a crucial point. The function $X(t)$ is a free function of time. We could be leaving $X$ to be a constant, leading to an exact $\Lambda$CDM background. However, this choice would most probably give, for eMTMG, the same fit as the fit of the Supernovae data by GR. Hence, we want to go beyond this possibility. In principle, there are an infinite number of choices for $X(t)$ and here we will explore one of the simplest, namely the transition of $X(t)$ between two different constants. This means that the theory, at the background level, will mimic two $\Lambda$CDM models, the ones reached before and after the transition occurs.

We should emphasize again that each model of eMTMG is defined only after $X(t)$ is given. A different $X(t)$ will generate not only a new background but also different dynamics for the perturbations, namely a different eMTMG model. Similarly, a form of $X(t)$ parametrized by a set of parameters define a subclass of eMTMG. Hence, eMTMG actually corresponds to a plethora of models, or in other words, to a framework for modified gravity models.

Therefore, let us consider the following parametrized form of $X$. 
\begin{equation}
X=X(z)=1+\frac{(A_{1}-1)\left[1+\tanh\!\left(\frac{A_{2}-z}{A_{2}A_{3}}\right)\right]}{1+\tanh\!\left(A_{3}^{-1}\right)}\,,
\end{equation}
where $z=a_0/a-1$ is the redshift, and we assume $A_{2}>0$ and $A_{3}>0$. If $A_1=1$ then $X$ remains constant during the evolution of the universe and this leads to a background that is identical to the one of $\Lambda$CDM. According to the chosen profile of $X$, we have that
\begin{align}
X_{{\rm -\infty}} & \equiv X(z=-\infty) =1+\frac{2(A_{1}-1)}{1+\tanh\!\left(A_{3}^{-1}\right)}\,,\\
X_{{\rm +\infty}} & \equiv X(z=+\infty) =1\,.
\end{align}
We impose that in the far future ($z\to-\infty$)
\begin{equation}
X_{{\rm -\infty}}>0\,,\quad{\rm or}\quad A_{1}>\frac{1}{2}\,[1-\tanh(A_{3}^{-1})]>0\,.
\end{equation}
Also, we have, without approximations, that
\begin{equation}
X_{0}\equiv X(z=0)=A_{1}>0\,.
\end{equation}
Also we have, since $\mathcal{N}=\ln(a/a_{0})$, that
\begin{equation}
X'=\frac{dX}{d\mathcal{N}}=-(1+z)\,\frac{dX}{dz}=(A_{1}-1)\,\frac{(1+z)\left[1-\tanh^{2}\left(\frac{A_{2}-z}{A_{2}A_{3}}\right)\right]}{A_{2}A_{3}[1+\tanh(A_{3}^{-1})]}\,,
\end{equation}
leading to
\begin{equation}
X'<0\,,\quad{\rm implies}\quad0<A_{1}<1\,.
\end{equation}
Therefore if $0<A_{1}<1$, then $X$ monotonically decreases from
unity (at high redshifts) to $A_{1}$  (at $z=0$).

The reason why we can set $X_{{\rm +\infty}}$ to unity without loss of generality, is that we can re-scale the coefficients such that this is possible. Suppose, for
instance, that we have $X_{{\rm +\infty}}\neq1$ but $\lim_{z\to+\infty}X(z)=X_{{\rm +\infty}}$,
with $0<X_{{\rm +\infty}}<+\infty$, then we would have, as shown in Eq. \eqref{eq:Frd1}, that the effective energy density of this eMTMG model can be written in general as
\begin{equation}
\rho_{X}=m^{2}M_{\rm P}^{2}\,(c_{4}-6\xi^{2}X^{3})\,.
\end{equation}
Then we can redefine the parameters
\begin{equation}
c_{4}=\bar{c}_{4}\,\frac{H_{0}^{2}}{m^{2}}\,,\quad\xi=\bar{\xi}\,\frac{H_{0}}{m}\,X_{{\rm +\infty}}^{-3/2}\,,
\end{equation}
 such that
\begin{equation}
\rho_{X}=M_{\rm P}^{2}H_{0}^{2}\,[\bar{c}_{4}-6\bar{\xi}^{2}(X/X_{{\rm +\infty}})^{3}]\,,
\end{equation}
effectively re-scaling $X$ to $\bar{X}\equiv X/X_{{\rm +\infty}}$. A similar procedure can be extended to all other eMTMG parameters, making it possible to freely impose that $X_{+\infty}=1$.

Let us then re-consider the Friedmann equation:
\begin{equation}
3M_{\rm P}^{2}H^{2}=\sum_{I}\rho_{I}+\rho_{X}\,,
\end{equation}
where the sum over $I$ runs over the known matter species. Then we
can also redefine
\begin{equation}
\varrho_{I}=\frac{\rho_{I}}{3M_{\rm P}^{2}}\,,\quad\varrho_{X}=\frac{\rho_{X}}{3M_{\rm P}^{2}}=\frac{1}{3}\,H_{0}^{2}\,[\bar{c}_{4}-6\bar{\xi}^{2}X^{3}]\,,\label{eq:rhoX}
\end{equation}
where we have once for all set $X_{{\rm +\infty}}$ to unity, so that
\begin{equation}
\frac{H^{2}}{H_{0}^{2}}=\frac{1}{H_{0}^{2}}\sum_I\varrho_{I}+\frac{1}{3}\,(\bar{c}_{4}-6\bar{\xi}^{2}X^{3})\,.
\end{equation}
By calling $\Omega_I=\varrho_I(z=0)/H_0^2$, today we have
\begin{equation}
\Omega_{{\rm DE},0}\equiv1-\sum_{I}\Omega_{I}=\frac{1}{3}\,(\bar{c}_{4}-6\bar{\xi}^{2}A_{1}^{3})\,,\quad{\rm or}\quad\bar{c}_{4}=3\Omega_{{\rm DE},0}+6\bar{\xi}^{2}A_{1}^{3}\,.
\end{equation}
On combining the Friedmann equation with the conservation equations
for each fluid $\rho_{I}$, one derives the following expression for
the pressure of the $X$ component
\begin{equation}
p_{X}\equiv\frac{P_{X}}{3M_{\rm P}^{2}}=-2H_{0}^{2}\bar{\xi}^{2}X^{2}(1+z)\,\frac{dX}{dz}-\varrho_{X}\,.
\end{equation}

Because of Eq.\ \eqref{eq:constr}, for the variable $r$ to be always a positive quantity, we need to impose that
$X+\frac{dX}{d\mathcal{N}}>0$. If $A_{1}\geq1$ this relation is
trivially satisfied, that is
\begin{equation}
A_{1}\geq1,\quad A_{2}>0\,,\quad A_{3}>0\,.
\end{equation}
Instead, if $0<A_{1}<1$, we proceed as follows. First of all, we
consider the function
\begin{equation}
\mathcal{F}(z)\equiv X-(1+z)\,\frac{dX}{dz}\,.
\end{equation}
and we require that $\mathcal{F}(z)>0$, $\forall z\geq0$. For $z>0$, we notice that its derivative vanishes for $z=A_{2}>0$, i.e.\ $\mathcal{F}_{,z}(A_2)=0$. Since, for $|\epsilon|\ll1$, $\mathcal{F}(A_2+\epsilon)\approx\mathcal{F}(A_2)+A_2^{-3}A_3^{-3}(1-A_1)(1+A_2)\epsilon^2/(1+\tanh{A_3^{-1}})$, then for $0< A_1<1$ the function $\mathcal{F}$ has a minimum in $z=A_2$. Then we have that
\begin{equation}
\mathcal{F}(z=A_{2})=0\,,\quad{\rm implies}\quad A_{1}=A_{1,{\rm min}}\equiv\frac{1+A_{2}-A_{2}A_{3}\tanh(A_{3}^{-1})}{1+A_{2}+A_{2}A_{3}}\,.
\end{equation}
If we consider $0<A_{3}<1$, and $A_{2}>0$, then $A_{1,{\rm min}}>0$,
always. Furthermore, in the same range of parameters, we also see
that $A_{1,{\rm min}}<1$. Then we need to assume $A_{1}>A_{1,{\rm min}}$ to impose $r(z)>0$ at all redshifts (times).
We can then define a convenient variable $\Delta_{1}$ so that the parameters of the background for the theory satisfy
\begin{alignat}{4}
A_{1}&=A_{1,{\rm min}}+\Delta_{1}\,,\quad&\Delta_{1}&>0\,,\\
A_2 &>0\,,&0<A_3&<1\,.
\end{alignat}
Here, we have disregarded the region $A_3\geq 1$ since the observational data disfavor such a slow transition. Indeed, the upper limit on $A_3$ does not reach $A_3=1$, as one can see in Table I. 

%%%%%%
%%%%%%
%%%%%%

\section{Linear cosmological perturbation theory}\label{sec:perturbations}

Although the background seems to be quite simple, the dynamics of the cosmological linear perturbation fields need to be investigated in detail. First of all, the constraints of the theory remove all the additional degrees of freedom leaving only the massive tensor modes to propagate in the gravitational sector. The vector modes are instead exactly equal to the GR counterpart. However, in the scalar sector, although eMTMG does not add any new degree of freedom compared to GR, we find substantial differences in the dynamics of the perturbation fields, as, for instance, $G_{\rm eff}/G_N$ differs from unity at late times. Therefore the largest constraints on the parameters of the theory will exactly come from all the constraints arising from the scalar sector.

\subsection{Scalar perturbation equations of motion}

In the following, we will consider the dynamics of all the perturbation variables. First of all the perturbations of the gravitational sector. Namely
\begin{align}
    N&=N(t)\,(1+\alpha)\,,\\
    N_i\, \mathrm{d} x^i&= N(t)\partial_i\chi\,\mathrm{d} x^i\,,\\
    \gamma_{ij} &= a(t)^2\,(1+2\Phi)\delta_{ij}+2\partial_i\partial_j E\,.
\end{align}
Furthermore, we need to consider the perturbation fields relative to the Lagrange multipliers as
\begin{align}
    \lambda&=\delta\lambda\,,\\
    \lambda^i&=\frac{\delta^{ij}}{a^2}\,\partial_j\lambda_V\,.
\end{align}

In the unitary gauge, the fiducial metric and lapse do not have any contributions from the perturbations, such that we still have ${\tilde\gamma}_{ij}=\tilde a^2\delta_{ij}$, and $M=M(t)$. As a consequence, the tensor ${\bar\zeta}^i{}_j$ still has the same form of Eq.\ \eqref{eq:zeta_ij}.

We treat matter fields as composed of perfect fluids with energy density $\rho_I$, where $I$ is the label for the fluid itself comprising baryons, cold dark matter, radiation gas, etc. Then we introduce the Schutz-Sorkin action \cite{Schutz:1977df,DeFelice:2015moy} for each $I$-th perfect fluid as
\begin{equation}
    S_{I}=-\int d^4x N\sqrt\gamma\left[ 
    \rho_I(n_I)+ J_I^0\partial_t\ell_I
    + J_I^i D_i\ell_I
    \right],
\end{equation}
where
\begin{equation}
    n_I\equiv\sqrt{-[-(N^2-N_iN^i)(J_I^0)^2+2J_I^0J_I^iN_i+\gamma_{ij}J_I^iJ_I^j]}\,,
\end{equation}
represents the number density of the $I$-th species. This action implies that for each matter species, we have the following perturbation variables
\begin{align}
    J^0_I&=\frac{J_I(t)}{N(t)}\,(1+\delta \mathcal{J}_I)\,,\\
    J^i_I&=\frac{\delta^{ik}}{a^2}\partial_k\,\delta\mathcal{J}_{I}^V\,,\\
    \ell_I&=\ell_I(t)+\delta\ell_I\,,
\end{align}
so that, at the lowest order, we find $n_I=J_I(t)$. If we expand the action at first for the field $\delta\ell_I$, then we find a background equation of motion that leads to
\begin{equation}
    n_I(t)=J_I(t)=\frac{\bar{\mathcal{N}}_{I}}{a^3}\,,
\end{equation}
which, being $\bar{\mathcal{N}}_{I}$ a constant, states that the number density is proportional to $a^{-3}$, as expected for a perfect fluid component. The background equation of motion for the field $\delta \mathcal{J}_I$ leads instead to a second background equation of motion which is satisfied by the following solution
\begin{equation}
    \ell_I(t)=-\int^tN(t')\,\rho_{I,n_I}\!(t')\,dt'\,,
\end{equation}
where $\rho_{I,n_I}=\partial\rho_I/\partial n_I$. As already mentioned earlier, the constraints of the theory impose the condition $\lambda(t)=0$ and then every other background consideration goes back to the results of the previous section. We are now ready to move into discussing the dynamics of the perturbation's equations of motion. To achieve this goal, we  expand the action in second order for the perturbation fields.

We can make some convenient field redefinitions. For instance, for each perfect fluid component, we have that the four-velocity $u_I^\mu$ is related to the four-vector $J_I^\mu$ by the relation $u_I^\mu = J_I^\mu/n_I$. Hence, we can define
\begin{equation}
    u_{Ii} = g_{i\beta} u^\beta = \frac1{n_I}\,[g_{i0} J_I^0 + g_{ik} J_I^k]\,, 
\end{equation}
where $g_{\alpha\beta}$ is the four-dimensional metric, built out of the ADM variables. In other words, focusing only on the scalar sector of the perturbations, we can introduce the velocity field $v_I$ as
\begin{equation}
    \partial_i v_I = u_{Ii} = \frac1{n_I}[J_I^0\, N_i + \gamma_{ik} J_I^k]\,.
\end{equation}
Out of this definition, we find that we can make the following field redefinition for each matter component
\begin{equation}
    \delta\mathcal{J}_{I}^V = n_I(t)\,(v_I-\chi)\,.
\end{equation}

After this field redefinition, the field $v_I$ inside the Lagrangian is a Lagrangian multiplier. Its equation of motion can be solved, for each matter component, in terms of the field $\delta\ell_I$ as
\begin{equation}
    \delta\ell_I = \rho_{I,n_I}\!(t)\,v_I\,.
\end{equation}

We can introduce for each matter component the energy density perturbation as follows. First of all, we can define the field
\begin{equation}
    \frac{\delta\rho_I}{\rho_I} \equiv \frac{\rho_I(n_I)}{\rho_I(t)}-1\,.
\end{equation}
By expanding the numerator of the previous expression up to linear order in the perturbations we obtain
\begin{equation}
    \delta\mathcal{J}_{I} = \frac{\rho_I(t)}{n_I(t)\,\rho_{I,n_I}(t)}\,\frac{\delta\rho_I}{\rho_I}-\alpha\,,
\end{equation}
and we can use the previous relation as a field redefinition from $\delta\mathcal{J}_I$ to the perturbation field $\frac{\delta\rho_I}{\rho_I}$ for each matter species. In this way, we have replaced the native Schutz-Sorkin perturbation variables with new variables, $v_I$, and $\delta\rho_I/\rho_I$, that we find more convenient and whose interpretation is easier to understand. At this level, on following a procedure introduced in \cite{DeFelice:2020cpt}, we can add a term into the Lagrangian density $L$ of the perturbations to take care of the linear shear that might be present in matter fields as follows
\begin{equation}
    L \to L + N \,a^{3} \sigma_{I} \left(
    \rho_{I} \frac{\delta\rho_I}{\rho_I}
    +3 n_{I} \rho_{I,n_I} \Phi \right).
\end{equation}
We can still use the freedom of choosing the time variable to set $N(t)=a(t)$ to be consistent with the conventions used in the code (based on CLASS) that we will employ in the numerical analysis. We also perform the following field redefinition to make use of the gauge invariant Bardeen potentials $\phi$ and $\psi$ and to match other gauge-invariant fields present in the CLASS code as follows
\begin{align}
    \alpha &=\psi-\frac{\dot\chi}{a}
    +\frac{1}{a}\,\frac{d}{dt}\!\left[ a\frac{d}{dt}\!\left(\frac{E}{a^2} \right)\right],\\
    \Phi &= -\phi -H\chi +aH \frac{d}{dt}\!\left(\frac{E}{a^2} \right),\\
    \frac{\delta\rho_I}{\rho_I}&=\delta_I
        -\frac{\dot{\rho}_I}{a\rho_I}\,\chi
        +\frac{\dot{\rho}_I}{\rho_I}\,\frac{d}{dt}\!\left(\frac{E}{a^2} \right),\\
    v_I &= -\frac{a}{k^2}\,\theta_I+\chi -a\,\frac{d}{dt}\!\left(\frac{E}{a^2} \right).
\end{align}
We have performed these field redefinitions but we still need to reduce the equations of motion to match the structure of the equations of motion present in GR. This is possible as the local physical degrees of freedom of eMTMG are the same as GR, although the equations for them change in the gravity sector. For this aim, we solve the equation of motion for the field $\lambda_V$ for the field $\chi$. We still need to solve for the remaining scalars $\lambda_V$, $\delta\lambda$, and $E$. The equation of motion for $\chi$ can be solved for $\delta\lambda$. The equation of motion for $\alpha$ can be solved for $E$. We can finally solve the shear equation, a linear combination of the equations of motion for $\Phi$ and $E$, for $\lambda_V$. At this stage, we have only the matter field scalar perturbations together with $\phi$ and $\psi$, as we wanted. The equations of motion for the matter perturbations exactly coincide with the ones of GR, since the Lagrangians of matter fields have not been modified. We now need two equations that will fix the two Bardeen potentials. One of these equations is the equation of motion for $E$ which sets a relation between $\dot\phi$ and all the other fields. This equation can be schematically written as
\begin{equation}
    \mathcal{E}_1 \equiv S_1\dot\phi +S_2\phi + S_3\psi +\sum_I S_I\,\theta_I = 0\,,
\end{equation}
where the coefficients $S$'s are functions of the wave number and of the background quantities, and we have labeled each matter component by an $I$. This equation can be used as a differential equation for $\phi$. The equation of motion for $\delta\lambda$ can be instead schematically written as
\begin{equation}
    \mathcal{E}_{\delta\lambda}\equiv U_1\phi + \sum_I U_I\delta_I + \sum_I \tilde{U}_I\theta_I = 0\,.
\end{equation}

The second equation of motion that we need to find for the variable $\psi$ is a linear combination of the equation of motion $\mathcal{E}_{\delta\lambda}$, its derivative, $\mathcal{E}_1$, and the matter equations of motion. The coefficients of this linear combination, except for the one multiplying $\mathcal{E}_{\delta\lambda}$, have been chosen to remove any time derivative of the fields. At this stage, this new equation can be solved algebraically for the field $\psi$ (this equation would correspond in GR to the shear equation) and can be written schematically as
\begin{equation}
    \mathcal{E}_2\equiv V_1\psi +V_2\phi +\sum_I V_I\delta_I + \sum_I \tilde{V}_I \theta_I + \sum_I \bar{V}_I\sigma_I = 0\,.
\end{equation}
Once we solve it for $\psi$, we can check that the coefficient of $\phi$ tends to unity at early times, whereas the coefficients $(V_I,\tilde{V}_I)\to(0,0)$ in the limit $m/H\to0$, i.e.\ at early times. This behavior was fixed by properly choosing the coefficient of  $\mathcal{E}_{\delta\lambda}$ in the linear combination defining $\mathcal{E}_2$. We also mention here that each of the coefficients $V_I$ contains $\partial^2\rho_I/\partial n_I^2$ (which is then related to the speed of propagation of the $I$-th matter field, ${c_{s}^2}_I=n_I \rho_{I,n_I,n_I}/\rho_{I,n_I}$), and can be schematically written as $V_I=V^{(A)}{c_{s}^2}_I\rho_I + V^{(B)}\rho_I$, where now both $V^{(A)}$ and $V^{(B)}$ are independent of the specific $I$-th matter field quantities. Finally, the term proportional to the shear term is the same as GR. Now, the CLASS code can be easily modified to solve the new equations of motion of eMTMG.

\subsection{Vector modes and gravitational waves}
The theory does not add any local physical degrees of freedom in the vector sector by construction and the equations of motion do not get modified from GR. This happens because the constraint due to $\lambda^i$ sets the vector components of $\delta\gamma_{ij}$ to vanish identically. In GR, the effect of this equation of motion corresponds to setting a flat gauge for the vector modes, as $\delta\gamma_{ij}=a\,(\partial_i C^V_j+ \partial_j C^V_i)$, where $\delta^{ij}\partial_i C^V_i = 0$, and the flat gauge corresponds to setting $C^V_i=0$. In other words, the vector components of the field $\lambda^i$ set the constraints $C^V_i=0$, which, in turn, makes the dynamics of the vector modes identical to the one of GR.

As for the tensor modes, the metric tensor is now defined as 
\begin{equation}
    \gamma_{ij} = a^2\,(\delta_{ij} + h_{ij})\,,
\end{equation}
where the traceless and transverse conditions hold, namely $\delta^{ij}h_{ij} = 0$ and $\delta^{ik}\partial_k h_{ij}=0$. By using these properties and by decomposing into two polarizations as $h_{ij} = \sum_{\lambda={+},{\times}}\epsilon^\lambda_{ij} h_\lambda$, the two polarization tensors being normalized to unity and orthogonal to each other, we find that the Lagrangian density for the tensor modes can be written as
\begin{equation}
    L_{\rm GW}=\frac{M_{\rm P}^2}{8}\,Na^3\sum_\lambda\left[
    \frac{\dot{h}_\lambda^2}{N^2} - \frac{(\partial h_\lambda)^2}{a^2}-\mu^2\,h_\lambda^2
    \right],
\end{equation}
where
\begin{equation}
    \mu^2 = 6m^2 r\,\bigl(\zeta_2^2 +2\zeta_3^2/X + \tfrac95\,\zeta_4^2/X^2\bigr)\,.\label{eq:mass_GW}
\end{equation}
Therefore the theory introduces a non-vanishing mass into the tensor sector of the theory, allowing FLRW solutions to exist. This is the main motivation for introducing eMTMG as an alternative to the standard lore of a scalar field-driven dark energy scenario. This feature together with the theory being minimal, i.e.\ without having extra local physical degrees of freedom (which might be unstable), and the property of having an always non-zero squared mass for the tensor mode and a finite $G_{\rm eff}/G_N$ at all times. Understanding the value of the mass of these tensor modes, via an analysis of the most recent data available, corresponds to the goal of our study. We are now ready to fit the data with the dynamics determined by eMTMG for the background and the perturbation variables. 

\section{Data and Methodology}
\label{data}

In the following, we attempt to explore the response of observational data to the free parameters of the simplest model within the eMTMG framework, and the chosen background profile $X=X(t)$ for the theory.

We describe below the observational data sets and the statistical methods that we use to explore the parameter space of the theory.
\begin{itemize}
\item \textbf{CMB:} From the \textit{Planck} 2018 legacy data release, we use the CMB measurements, viz., high-$\ell$ \texttt{Plik} TT likelihood (in the multipole range $30 \leq \ell \leq 2508$), TE and EE (in the multipole range $30 \leq \ell \leq 1996$),  low-$\ell$ TT-only ($2 \leq \ell \leq 29$), the low-$\ell$ EE-only ($2 \leq \ell \leq 29$) likelihood \cite{Planck:2019nip_a}, in addition to the CMB lensing power spectrum measurements \cite{Planck:2018lbu_b}. We refer to this dataset as \texttt{Planck}.

\item \textbf{BAO:} From the latest compilation of  Baryon Acoustic Oscillation (BAO) distance and expansion rate measurements from the SDSS collaboration, we use 14 BAO measurements, viz., the isotropic BAO measurements of $D_V(z)/r_d$ (where $D_V(z)$ and $r_d$ are the spherically averaged volume distance, and sound horizon at baryon drag, respectively) and anisotropic BAO measurements of $D_M(z)/r_d$ and $D_H(z)/r_d$ (where $D_M(z)$ and $D_H(z)=c/H(z)$ are the comoving angular diameter distance and the Hubble distance, respectively), as compiled in Table 3 of \cite{Alam:2020sor}. We refer to this dataset as \texttt{BAO}.

\item \textbf{Type Ia supernovae:} We use the SNe Ia distance moduli measurements from the Pantheon+ sample \cite{Brout:2022vxf}, which consists of 1701 light curves of 1550 distinct SNe Ia ranging in the redshift interval $z \in [0.001, 2.26]$. We refer to this dataset as \texttt{PantheonPlus}. 

\item \textbf{Cosmic Shear:} We use KiDS-1000 data~\cite{Kuijken:2019gsa,Giblin:2020quj}. This includes the weak lensing two-point statistics data for both the auto and cross-correlations across five tomographic redshift bins~\cite{Hildebrandt:2020rno}. We also employ the KiDS-1000 public likelihood\footnote{\href{https://github.com/BStoelzner/KiDS-1000_MontePython_likelihood}{KiDS-1000 Montepython likelihood.}}. We follow the KiDS team analysis and adopt the COSEBIs (Complete Orthogonal Sets of E/B-Integrals) likelihood in our results~\cite{KiDS:2020suj}. For the prediction of the matter power spectrum, we use the augmented halo model code, HMcode~\cite{Mead_2015}. We highlight that at the level of the linear perturbations theory and Boltzmann equations, the model described here is well modeled, in the sense that the theory possesses a Lagrangian out of which one determines uniquely, once the background $X(t)$ is given, the dynamics of the system. This would still hold at non-linear scales, as in GR. For MTMG, a model that is very closely related to eMTMG, a study appeared, see \cite{Hagala:2020eax}, that consisted of an analysis of the gravitational N-body simulation for the theory in an expanding universe during dust domination. The analysis showed the problem of MTMG, already known at the level of linear perturbation theory, for which the effective gravitational constant could behave badly in under-dense regions. The theory discussed here eMTMG does not possess, by construction, such a problem, and in this case, we believe that the transition from a cosmological environment to the highly dense (or under-dense) one should be smooth. Furthermore, the mass of the graviton is bound to be of order $H_0$. This corresponds to an extremely low value which is still beyond the possibility of observation in LIGO. In particular, all the gravitational waves produced at astrophysical scales will have a spectrum peaked at energies much higher than $H_0$, making these waves practically ultra-relativistic. Since the more we enter the small and non-linear scales of galaxies, the more the environment energy is available to make the gravitational wave more and more ultra-relativistic, then, in this limit, eMTMG will reduce to GR. Therefore, we can use the same non-linear results found in GR also in eMTMG. We refer to this data set as \texttt{KiDS-1000}. 

\end{itemize}

This shows that we have chosen recent data and the ones that we believe will set strong constraints on the model parameters. The model baseline of eMTMG is given by 
\begin{equation*}
    \mathcal{P}= \left\{ \omega_{\rm b}, \, \omega_{\rm c}, \, \theta_s, \,  A_{\rm s}, \, n_{\rm s}, \, \tau_{\rm reio},
\,  \Delta_{1}, \, A_2, \, A_3, \, \xi, \, \zeta_1, \, \zeta_2, \, \zeta_3, \, \zeta_4 \right\}
\end{equation*}
where the first six are the common ones with $\Lambda$CDM, viz., $\omega_{\rm b}=\Omega_{\rm b} h^2$ and $\omega_{\rm c}=\Omega_{\rm c}h^2$ ($\Omega$ being the present-day density parameter) are, respectively, the present-day physical density parameters of baryons and CDM, $\theta_{\rm s}$ is the ratio of the sound horizon to the angular diameter distance at decoupling, $A_{\rm s}$ is the initial super-horizon amplitude of curvature perturbations at $k=0.05$ Mpc$^{-1}$, $n_{\rm s}$ is the primordial spectral index, and $\tau_{\rm reio}$ is the reionization optical depth. The parameters $\Delta_{1}, \, A_2, \, A_3, \, \xi, \, \zeta_1, \, \zeta_2, \, \zeta_3, \, \zeta_4$, are additional free parameters related to the eMTMG dynamics.

The first three eMTMG parameters $\Delta_{1}$, $A_2$, and $A_3$ are all parameters that are connected with the (free) choice of the background profile $X(t)$ for the theory. In particular, the choice of the $X(t)$ was made to have a dark component which was interpolating two values of the effective cosmological constant.  The non-negative parameter $\Delta_1$ corresponds to the difference between $A_1$ and its minimum value (that is also positive but smaller than unity). In turn, $A_1$ determines how different the model is from $\Lambda$CDM. If $A_1 =1$, the background exactly reduces to the one of $\Lambda$CDM, that is no transition occurs. If $A_1\neq1$ we will have in general the above-mentioned transition between two different effective cosmological constants. In particular, on examining Eq.\ \eqref{eq:rhoX}, we find that 
$\rho_{X}(z=+\infty)-\rho_{X}(z=0)=6M_{\rm P}^2 H_0^2\xi^2\,(A_1^3-1)$. Therefore a value for $A_1$ larger than unity will lead to an effective cosmological constant larger in the past. The parameter $A_2$ determines instead the redshift of transition, whereas the parameter $A_3$ determines the width of transition in redshift, or how fast the transition is. 

Besides the choice of the background evolution, we still need to give the parameters intrinsic to the theory, namely $\xi$, $\zeta_i$ ($i=\{1,\dots,4\}$) that determine, together with the chosen background, the dynamics of the observables to be analyzed.

We use the Metropolis-Hastings model in an appropriately modified version of \texttt{CLASS}+\texttt{MontePython} code \cite{Blas:2011rf, Audren:2012wb, Brinckmann:2018cvx}  to derive the constraints on cosmological parameters using various data combinations from the data sets described above, ensuring a Gelman-Rubin convergence criterion of $R - 1 < 10^{-2}$. In what follows, we describe our main results.

\begin{table*}
\centering
\label{tab:table_cmb}
\caption{Marginalized constraints (mean values with 68\% CL) on the free and some derived parameters of the eMTMG model for different data set combinations. For some parameters, 95\% CL upper bounds are given. }
\begin{tabular}{|l|l|l|l|l|}
\hline
Parameter & Planck  & Planck + PantheonPlus & Planck + PantheonPlus + BAO \\
\hline
$10^{2}\omega{}_{b}$ & $2.239^{+0.016}_{-0.013}$ & $2.235\pm 0.013$ & $2.237\pm 0.012$ \\

$\omega{}_{\rm cdm}$ & $0.1200\pm 0.0012$ & $0.1204\pm 0.0011$ & $0.12030\pm 0.00087$ \\

$\ln10^{10}A_{s}$ & $3.047\pm 0.015$ & $3.042\pm 0.014$ & $3.043\pm 0.012$ \\

$n_{s}$ & $0.9651^{+0.0041}_{-0.0046}$ & $0.9644\pm 0.0039$ & $0.9647\pm 0.0034$ \\

$\tau{}_{\rm reio}$ & $0.0550\pm 0.0076$ & $0.0524\pm 0.0073$ & $0.0536\pm 0.0060$ \\

$H_0$ [km/s/Mpc] & $67.96\pm 0.54$ & $67.70\pm 0.49$ & $67.81\pm 0.40$ \\

$\Omega{}_{m}$ & $0.3085\pm 0.0073$ & $0.3114\pm 0.0066$ & $0.3104\pm 0.0053$ \\

$S_8$ & $0.836\pm 0.014$ & $0.839\pm 0.012$ & $0.838\pm 0.010$ \\

$\theta_0$ & $< 0.739$ & $< 0.463$ & $< 0.469$ \\

$\Delta_1$ & $< 4.65$ & $< 7.22$ & $< 3.06$ \\

$A_1$      & $1.72^{+0.17}_{-1.5}$ &  $2.58^{+0.28}_{-2.3}$    & $1.33^{+0.33}_{-1.1}$  \\

$A_2$ & $< 94.5$ & $< 84.6$ & $< 92.8$ \\

$A_3$ & $0.49^{+0.29}_{-0.14}$ & $0.46^{+0.33}_{-0.42}$ & $0.50^{+0.27}_{-0.13}$ \\

$\xi$ & $< 1.08$ & $< 0.257$ & $< 0.507$ \\

$\zeta_1$ & $< 444$ & $< 53.9$ & $1226^{+500}_{-900}$ \\

$\zeta_2$ & $525\pm 200$ & $< 164$ & $10840^{+5000}_{-7000}$ \\

$\zeta_3$ & $< 483$ & $< 64.7$ & $< 6160$ \\

$\zeta_4$ & $< 879$ & $< 53.1$ & $< 5980$ \\
\hline
\end{tabular}
\end{table*}

\begin{figure}[b!]
    \centering
    \includegraphics[width=8.cm]{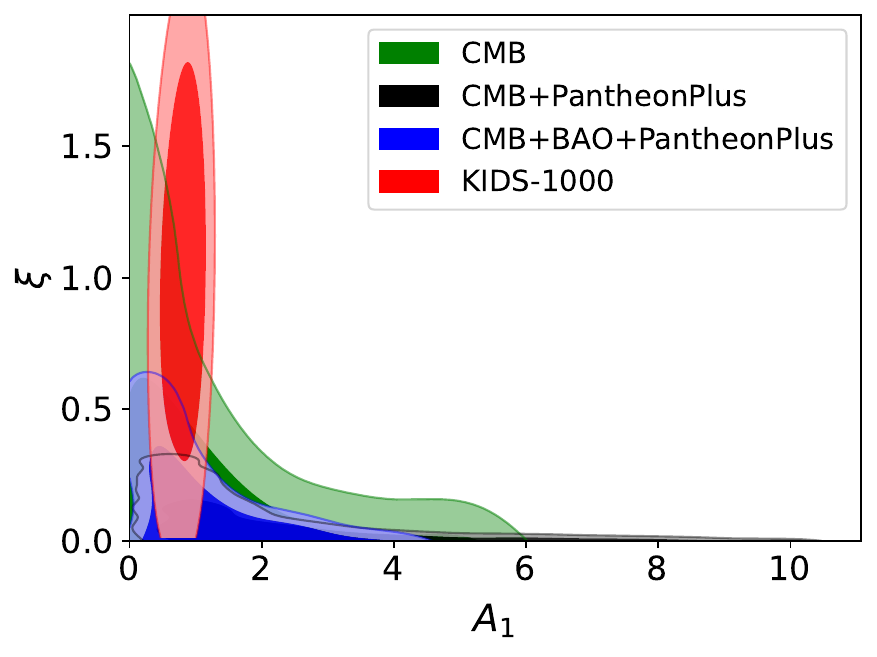}
    \caption{1D and 2D marginalized posterior probability distributions of eMTMG model parameters $A_1$ and $\xi$  obtained from the CMB, CMB+PantheonPlus, CMB + PantheonPlus + BAO and KIDS-1000 dataset combinations.}
    \label{fig:PS_model}
\end{figure}

\begin{figure}[b!]
    \centering
    \includegraphics[width=11.cm]{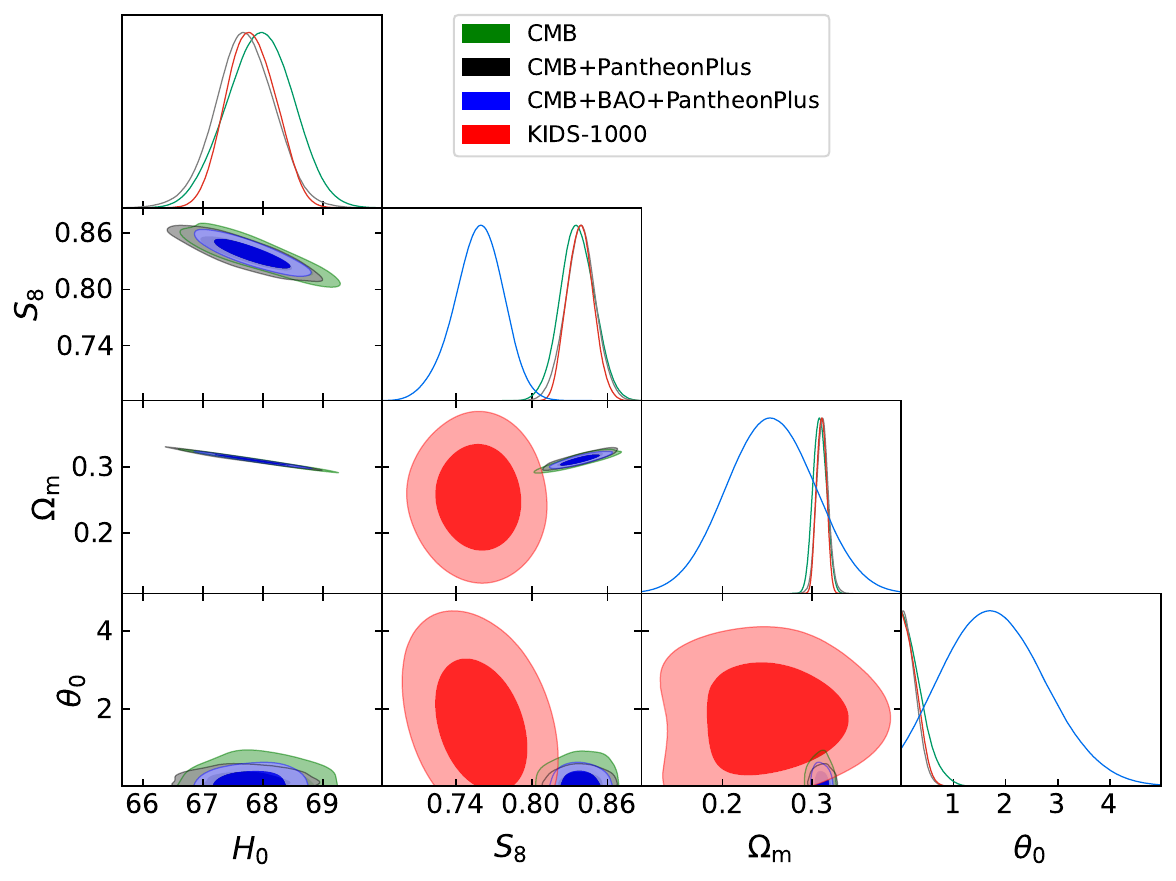}
    \caption{Triangular plot showing the 2D joint and 1D marginalized posterior probability distributions for $H_0$, $S_8$, $\Omega_m$ and $\theta_0$ obtained from the CMB, CMB+PantheonPlus, CMB + PantheonPlus + BAO and KIDS-1000 dataset combinations within the eMTMG model.}
    \label{fig:2}
\end{figure}

\section{Results}
\label{results}

Table I summarizes the main results of our statistical analyses from the Planck CMB dataset and its combination with PantheonPlus and BAO data.  Figure \ref{fig:PS_model} shows the parametric space at 68\% and 95\% CL of  two main free parameters of the eMTMG model, from the 4 analyses carried out in this work. Figure \ref{fig:2} shows the parametric space on the baseline $H_0$, $S_8$, $\Omega_m$ and $\theta_0$. We begin by examining the constraints on the full parametric space of the model using Planck data only.  At the level of the free and extra parameters of the theory, we find $\theta_0 = \mu(t=0)/H_0 < 0.79$ at 95\% CL, which bounds the graviton mass to be $ < 1.08 \times 10^{-33}$ eV at  95\% CL. We note $A_1=1$, the value corresponding to the $\Lambda$CDM background, is well inside the 1-$\sigma$ contour. Thus, we only have an upper and lower limit for $A_1$ that does not discard the dynamics of the model beyond the $\Lambda$CDM model at the background level. On the other hand, the effective dark energy behavior induced by corrections of the model presented here is quantified by the function $X(z)$, which depends also on other two parameters $A_2$, and $A_3$, as introduced previously. From the Planck analysis, we find only upper limits on $A_2$, while $A_3 = 0.49 \pm 0.32$ at 68\% CL. All these constraints on the free parameters of the model fix an upper limit that does not discard the dynamics of the model beyond the $\Lambda$CDM model but shows that the eMTMG is statistically compatible with $\Lambda$CDM.
Now, returning to the interpretation of the baseline parameters common with the $\Lambda$CDM model, we find the that six common parameters between eMTMG and $\Lambda$CDM do not differ statistically. In particular, we note $H_0 = 68 \pm 1$ km/s/Mpc and $S_8 = 0.823 \pm 0.026$ at 68\% CL. In light of the current $H_0$ tension, we can interpret that the eMTMG model, with the chosen background $X(t)$ which makes a transition between two $\Lambda$CDM models, can only assuage this tension, but is not capable of giving a satisfactory solution on the tension.

As a follow-up analysis, we performed the joint analyses using Planck + PantheonPlus and Planck + PantheonPlus + BAO. We note an improvement in the bound of the graviton mass, now constrained $\theta_0 < 0.46$ at 95\% CL for both joint analyses. We find $\bar{\xi} <$ 0.26 (0.50) from Planck  + PantheonPlus (Planck  + PantheonPlus + BAO), respectively. Significant improvement in the other parameters of the theory is not observed. For the main common parameters between eMTMG and $\Lambda$CDM, as expected, we note a natural improvement in the accuracy of all parameters when combined with PantheonPlus and BAO, as may be seen in Table I. As the data used in this work are the most recent and robust geometrical samples available up-to-date in the literature, it is not expected that additional geometrical distance data can improve the accuracy of our model baseline parameters.

As the eMTMG predicts a non-trivial effective gravitational constant, which can influence the evolution of the matter density field fluctuations and large-scale structure observable, we also analyze the model in light of the KiDS-1000 samples. We first note that the KIDS-1000 dataset only weakly constrains the graviton mass, as we find $\theta_0 < 3.38$ at 95\% CL. There is an order of magnitude in differences when compared with CMB and CMB plus geometrical data. Also for these data, there is no evidence for a background difference from $\Lambda$CDM ($A_1=1$ case) as $A_1 = 0.81^{+0.27}_{-0.38}$ at 68\% CL. However, it is interesting to note that, in the case of transition, we have $A_2 =3.98^{+0.40}_{-0.26}$ at 68\% CL, which indicates evidence that the transition happens, if present, at high redshifts, $z \sim 4$.

We can note that Planck, Planck + PantheonPlus, and Planck + PantheonPlus + BAO, are in tension with KIDS-1000 only analysis. This tension is also clearly transferred to the mass of the graviton, quantified by $\theta_0$. This tension in the physical parameters of the eMTMG can be interpreted as a result of the tension in $S_8$, where the $S_8$ tension is being transferred statistically to other parameters.

\section{Final remarks}
\label{final}

Modifications of general relativity have been an extended research line of investigation in the last two decades in cosmology, primarily motivated to explain the physical mechanism behind the accelerated expansion of the Universe at late time. In particular, models that give rise to a mass for the gravitational waves have attracted interest, as indeed the present acceleration of the universe could be the reason why the graviton is massive. It turns out that to build a model for a massive graviton that is at the same time free from ghost instabilities and endowed with viable and phenomenologically interesting cosmological background solutions is not an easy task. Recently, minimal models of massive gravity have given the chance to achieve the goal of having a theory of a massive graviton, which has the possibility of explaining the cosmological evolution of our universe. In light of these modes, the one dubbed as ``eMTMG'' provides a framework on which one can study the effects of a massive graviton on cosmological backgrounds. In particular, the model is built to have non-singular dynamics for the cosmological perturbations but still allows e.g.\ $0<G_{\rm eff}/G_N<1$ today. On top of that, eMTMG models have to be defined by means of a free background function, which determines the background dynamics. In this regard, this choice is analog to one of the potential for a quintessence theory, however, the model only possesses two gravitational degrees of freedom. The choice of the background function induces a change of $H(z)$ and the dynamics of the cosmological perturbation fields. This feature together with the property that $G_{\rm eff}/G_N<1$ today can in principle be used to address both the $H_0$ and $S_8$ tensions.

In this work, we present for the first time some observational constraints on the eMTMG \cite{DeFelice:2022mcd}, focusing on its particular subclass in which the effective cosmological constant experiences a transition from one value to another, as well as to impose an upper bound on the graviton mass in this subclass. This simple choice for the free background function was done in order to explore the dependence of the $H_0$ and $S_8$ parameters on it. We find that a simple transition between two different de Sitter vacua is not enough to address the above-mentioned tensions. This implies that in the playground of the possible choices for $H(z)$, to solve the tensions, a deeper search in function space is needed. On the other hand, we have also found that nowadays cosmological data are strong enough to set strong constraints on the mass of the graviton. In fact, although these constraints are model dependent, and in particular dependent on the choice of the free background function, still it shows that the data are able to give a stringent bound on the free parameters of eMTMG. In this sense, we find $\theta_0 = \mu(t=0)/H_0 < 0.46$ at 95 \% CL from the joint analysis Planck + PantheonPlus + BAO, which means an upper bound $< 6.6 \times 10^{-34}$ eV on the mass of the graviton. The observational constraints on the eMTMG derivatives here are consistent with a $\Lambda$CDM dynamics but given the great potential of the theory to provide several other dynamical characters for the function $X(z)$, as well as to give a non-trivial phenomenology for the growth of structure through the effective gravitational constant behavior from Eq.\ (\ref{Geff}). In future works, we hope to develop results that allow the theory to be able to explain the current existing anomalies and tensions in some cosmological parameters.

\begin{acknowledgments}
\noindent  The work of ADF was supported by the Japan Society for the Promotion of Science Grants-in-Aid for Scientific Research No.\ 20K03969. SK gratefully acknowledges support from the Science and Engineering Research Board (SERB), Govt.\ of India (File No.~CRG/2021/004658).
The work of SM was supported in part by the World Premier International Research Center Initiative (WPI), MEXT, Japan. SM is grateful for the hospitality of Perimeter Institute, the cosmology group at Simon Fraser University and the Theoretical Physics Institute at University of Alberta, where part of this work was carried out. Research at Perimeter Institute is supported in part by the Government of Canada through the Department of Innovation, Science and Economic Development and by the Province of Ontario through the Ministry of Colleges and Universities. 
RCN thanks the financial support from the Conselho Nacional de Desenvolvimento Cient\'{i}fico e Tecnologico (CNPq, National Council for Scientific and Technological Development) under project No.\ 304306/2022-3, and the Funda\c{c}\~{a}o de Amparo \`{a} pesquisa do Estado do RS (FAPERGS, Research Support Foundation of the State of RS) for partial financial support under project No.\ 23/2551-0000848-3.

\end{acknowledgments}

\bibliographystyle{mybibstyle}
\bibliography{extMTMG_biblio}

\begin{thebibliography}{59}%
\makeatletter
\providecommand \@ifxundefined [1]{%
 \@ifx{#1\undefined}
}%
\providecommand \@ifnum [1]{%
 \ifnum #1\expandafter \@firstoftwo
 \else \expandafter \@secondoftwo
 \fi
}%
\providecommand \@ifx [1]{%
 \ifx #1\expandafter \@firstoftwo
 \else \expandafter \@secondoftwo
 \fi
}%
\providecommand \natexlab [1]{#1}%
\providecommand \enquote  [1]{``#1''}%
\providecommand \bibnamefont  [1]{#1}%
\providecommand \bibfnamefont [1]{#1}%
\providecommand \citenamefont [1]{#1}%
\providecommand \href@noop [0]{\@secondoftwo}%
\providecommand \href [0]{\begingroup \@sanitize@url \@href}%
\providecommand \@href[1]{\@@startlink{#1}\@@href}%
\providecommand \@@href[1]{\endgroup#1\@@endlink}%
\providecommand \@sanitize@url [0]{\catcode `\\12\catcode `\$12\catcode
  `\&12\catcode `\#12\catcode `\^12\catcode `\_12\catcode `\%12\relax}%
\providecommand \@@startlink[1]{}%
\providecommand \@@endlink[0]{}%
\providecommand \url  [0]{\begingroup\@sanitize@url \@url }%
\providecommand \@url [1]{\endgroup\@href {#1}{\urlprefix }}%
\providecommand \urlprefix  [0]{URL }%
\providecommand \Eprint [0]{\href }%
\providecommand \doibase [0]{http://dx.doi.org/}%
\providecommand \selectlanguage [0]{\@gobble}%
\providecommand \bibinfo  [0]{\@secondoftwo}%
\providecommand \bibfield  [0]{\@secondoftwo}%
\providecommand \translation [1]{[#1]}%
\providecommand \BibitemOpen [0]{}%
\providecommand \bibitemStop [0]{}%
\providecommand \bibitemNoStop [0]{.\EOS\space}%
\providecommand \EOS [0]{\spacefactor3000\relax}%
\providecommand \BibitemShut  [1]{\csname bibitem#1\endcsname}%
\let\auto@bib@innerbib\@empty
%</preamble>
\bibitem [{\citenamefont {Perlmutter}\ \emph {et~al.}(1999)\citenamefont
  {Perlmutter}, \citenamefont {Aldering}, \citenamefont {Goldhaber},
  \citenamefont {Knop}, \citenamefont {Nugent}, \citenamefont {Castro},
  \citenamefont {Deustua}, \citenamefont {Fabbro}, \citenamefont {Goobar},
  \citenamefont {Groom},\ and\ \citenamefont {et~al.}}]{DE_obs_1}%
  \BibitemOpen
  \bibfield  {author} {\bibinfo {author} {\bibfnamefont {S.}~\bibnamefont
  {Perlmutter}}, \bibinfo {author} {\bibfnamefont {G.}~\bibnamefont
  {Aldering}}, \bibinfo {author} {\bibfnamefont {G.}~\bibnamefont {Goldhaber}},
  \bibinfo {author} {\bibfnamefont {R.~A.}\ \bibnamefont {Knop}}, \bibinfo
  {author} {\bibfnamefont {P.}~\bibnamefont {Nugent}}, \bibinfo {author}
  {\bibfnamefont {P.~G.}\ \bibnamefont {Castro}}, \bibinfo {author}
  {\bibfnamefont {S.}~\bibnamefont {Deustua}}, \bibinfo {author} {\bibfnamefont
  {S.}~\bibnamefont {Fabbro}}, \bibinfo {author} {\bibfnamefont
  {A.}~\bibnamefont {Goobar}}, \bibinfo {author} {\bibfnamefont {D.~E.}\
  \bibnamefont {Groom}},  and \bibinfo {author} {\bibnamefont {et~al.}},\
  }\href {\doibase 10.1086/307221} {\bibfield  {journal} {\bibinfo  {journal}
  {\emph {The Astrophysical Journal}}\ }\textbf {\bibinfo {volume} {517}},\
  \bibinfo {pages} {565} (\bibinfo {year} {1999})},\ \Eprint
  {http://arxiv.org/abs/astro-ph/9812133} {arXiv:astro-ph/9812133 [astro-ph]}
  \BibitemShut {NoStop}%
\bibitem [{\citenamefont {Riess}\ \emph {et~al.}(1998)\citenamefont {Riess},
  \citenamefont {Filippenko}, \citenamefont {Challis}, \citenamefont
  {Clocchiatti}, \citenamefont {Diercks}, \citenamefont {Garnavich},
  \citenamefont {Gilliland}, \citenamefont {Hogan}, \citenamefont {Jha},
  \citenamefont {Kirshner},\ and\ \citenamefont {et~al.}}]{DE_obs_2}%
  \BibitemOpen
  \bibfield  {author} {\bibinfo {author} {\bibfnamefont {A.~G.}\ \bibnamefont
  {Riess}}, \bibinfo {author} {\bibfnamefont {A.~V.}\ \bibnamefont
  {Filippenko}}, \bibinfo {author} {\bibfnamefont {P.}~\bibnamefont {Challis}},
  \bibinfo {author} {\bibfnamefont {A.}~\bibnamefont {Clocchiatti}}, \bibinfo
  {author} {\bibfnamefont {A.}~\bibnamefont {Diercks}}, \bibinfo {author}
  {\bibfnamefont {P.~M.}\ \bibnamefont {Garnavich}}, \bibinfo {author}
  {\bibfnamefont {R.~L.}\ \bibnamefont {Gilliland}}, \bibinfo {author}
  {\bibfnamefont {C.~J.}\ \bibnamefont {Hogan}}, \bibinfo {author}
  {\bibfnamefont {S.}~\bibnamefont {Jha}}, \bibinfo {author} {\bibfnamefont
  {R.~P.}\ \bibnamefont {Kirshner}},  and \bibinfo {author} {\bibnamefont
  {et~al.}},\ }\href {\doibase 10.1086/300499} {\bibfield  {journal} {\bibinfo
  {journal} {\emph {The Astronomical Journal}}\ }\textbf {\bibinfo {volume}
  {116}},\ \bibinfo {pages} {1009} (\bibinfo {year} {1998})},\ \Eprint
  {http://arxiv.org/abs/astro-ph/9805201} {arXiv:astro-ph/9805201 [astro-ph]}
  \BibitemShut {NoStop}%
\bibitem [{\citenamefont {Weinberg}(1989)}]{Weinberg:1988cp}%
  \BibitemOpen
  \bibfield  {author} {\bibinfo {author} {\bibfnamefont {S.}~\bibnamefont
  {Weinberg}},\ }\href {\doibase 10.1103/RevModPhys.61.1} {\bibfield  {journal}
  {\bibinfo  {journal} {\emph {Rev. Mod. Phys.}}\ }\textbf {\bibinfo {volume}
  {61}},\ \bibinfo {pages} {1} (\bibinfo {year} {1989})}\BibitemShut {NoStop}%
\bibitem [{\citenamefont {Padmanabhan}(2003)}]{Padmanabhan:2002ji}%
  \BibitemOpen
  \bibfield  {author} {\bibinfo {author} {\bibfnamefont {T.}~\bibnamefont
  {Padmanabhan}},\ }\href {\doibase 10.1016/S0370-1573(03)00120-0} {\bibfield
  {journal} {\bibinfo  {journal} {\emph {Phys. Rept.}}\ }\textbf {\bibinfo
  {volume} {380}},\ \bibinfo {pages} {235} (\bibinfo {year} {2003})},\ \Eprint
  {http://arxiv.org/abs/hep-th/0212290} {arXiv:hep-th/0212290} \BibitemShut
  {NoStop}%
\bibitem [{\citenamefont {Copeland}\ \emph {et~al.}(2006)\citenamefont
  {Copeland}, \citenamefont {Sami},\ and\ \citenamefont
  {Tsujikawa}}]{Copeland_2006}%
  \BibitemOpen
  \bibfield  {author} {\bibinfo {author} {\bibfnamefont {E.~J.}\ \bibnamefont
  {Copeland}}, \bibinfo {author} {\bibfnamefont {M.}~\bibnamefont {Sami}},  and
  \bibinfo {author} {\bibfnamefont {S.}~\bibnamefont {Tsujikawa}},\ }\href
  {\doibase 10.1142/S021827180600942X} {\bibfield  {journal} {\bibinfo
  {journal} {\emph {Int. J. Mod. Phys. D}}\ }\textbf {\bibinfo {volume} {15}},\
  \bibinfo {pages} {1753} (\bibinfo {year} {2006})},\ \Eprint
  {http://arxiv.org/abs/hep-th/0603057} {arXiv:hep-th/0603057} \BibitemShut
  {NoStop}%
\bibitem [{\citenamefont {Ishak}(2019)}]{Ishak:2018his}%
  \BibitemOpen
  \bibfield  {author} {\bibinfo {author} {\bibfnamefont {M.}~\bibnamefont
  {Ishak}},\ }\href {\doibase 10.1007/s41114-018-0017-4} {\bibfield  {journal}
  {\bibinfo  {journal} {\emph {Living Rev. Rel.}}\ }\textbf {\bibinfo {volume}
  {22}},\ \bibinfo {pages} {1} (\bibinfo {year} {2019})},\ \Eprint
  {http://arxiv.org/abs/1806.10122} {arXiv:1806.10122 [astro-ph.CO]}
  \BibitemShut {NoStop}%
\bibitem [{\citenamefont {Saridakis}\ \emph {et~al.}(2021)\citenamefont
  {Saridakis} \emph {et~al.}}]{CANTATA:2021ktz}%
  \BibitemOpen
  \bibfield  {author} {\bibinfo {author} {\bibfnamefont {E.~N.}\ \bibnamefont
  {Saridakis}} \emph {et~al.} (\bibinfo {collaboration} {CANTATA}),\ }\Eprint
  {http://arxiv.org/abs/2105.12582} {arXiv:2105.12582 [gr-qc]} \BibitemShut
  {NoStop}%
\bibitem [{\citenamefont {Heisenberg}(2019)}]{Heisenberg_2019}%
  \BibitemOpen
  \bibfield  {author} {\bibinfo {author} {\bibfnamefont {L.}~\bibnamefont
  {Heisenberg}},\ }\href {\doibase 10.1016/j.physrep.2018.11.006} {\bibfield
  {journal} {\bibinfo  {journal} {\emph {Physics Reports}}\ }\textbf {\bibinfo
  {volume} {796}},\ \bibinfo {pages} {1} (\bibinfo {year} {2019})}\BibitemShut
  {NoStop}%
\bibitem [{\citenamefont {Aghanim}\ \emph {et~al.}(2020)\citenamefont {Aghanim}
  \emph {et~al.}}]{Planck:2018vyg}%
  \BibitemOpen
  \bibfield  {author} {\bibinfo {author} {\bibfnamefont {N.}~\bibnamefont
  {Aghanim}} \emph {et~al.} (\bibinfo {collaboration} {Planck}),\ }\href
  {\doibase 10.1051/0004-6361/201833910} {\bibfield  {journal} {\bibinfo
  {journal} {\emph {Astron. Astrophys.}}\ }\textbf {\bibinfo {volume} {641}},\
  \bibinfo {pages} {A6} (\bibinfo {year} {2020})},\ \bibinfo {note} {[Erratum:
  Astron.Astrophys. 652, C4 (2021)]},\ \Eprint
  {http://arxiv.org/abs/1807.06209} {arXiv:1807.06209 [astro-ph.CO]}
  \BibitemShut {NoStop}%
\bibitem [{\citenamefont {Riess}\ \emph {et~al.}(2021)\citenamefont {Riess}
  \emph {et~al.}}]{Riess:2021jrx}%
  \BibitemOpen
  \bibfield  {author} {\bibinfo {author} {\bibfnamefont {A.~G.}\ \bibnamefont
  {Riess}} \emph {et~al.},\ }\Eprint {http://arxiv.org/abs/2112.04510}
  {arXiv:2112.04510 [astro-ph.CO]} \BibitemShut {NoStop}%
\bibitem [{\citenamefont {Riess}\ \emph {et~al.}(2022)\citenamefont {Riess},
  \citenamefont {Breuval}, \citenamefont {Yuan}, \citenamefont {Casertano},
  \citenamefont {Macri}, \citenamefont {Bowers}, \citenamefont {Scolnic},
  \citenamefont {Cantat-Gaudin}, \citenamefont {Anderson},\ and\ \citenamefont
  {Reyes}}]{Riess_2022}%
  \BibitemOpen
  \bibfield  {author} {\bibinfo {author} {\bibfnamefont {A.~G.}\ \bibnamefont
  {Riess}}, \bibinfo {author} {\bibfnamefont {L.}~\bibnamefont {Breuval}},
  \bibinfo {author} {\bibfnamefont {W.}~\bibnamefont {Yuan}}, \bibinfo {author}
  {\bibfnamefont {S.}~\bibnamefont {Casertano}}, \bibinfo {author}
  {\bibfnamefont {L.~M.}\ \bibnamefont {Macri}}, \bibinfo {author}
  {\bibfnamefont {J.~B.}\ \bibnamefont {Bowers}}, \bibinfo {author}
  {\bibfnamefont {D.}~\bibnamefont {Scolnic}}, \bibinfo {author} {\bibfnamefont
  {T.}~\bibnamefont {Cantat-Gaudin}}, \bibinfo {author} {\bibfnamefont {R.~I.}\
  \bibnamefont {Anderson}},  and \bibinfo {author} {\bibfnamefont {M.~C.}\
  \bibnamefont {Reyes}},\ }\href {\doibase 10.3847/1538-4357/ac8f24} {\bibfield
   {journal} {\bibinfo  {journal} {\emph {The Astrophysical Journal}}\ }\textbf
  {\bibinfo {volume} {938}},\ \bibinfo {pages} {36} (\bibinfo {year}
  {2022})}\BibitemShut {NoStop}%
\bibitem [{\citenamefont {Murakami}\ \emph {et~al.}(2023)\citenamefont
  {Murakami}, \citenamefont {Riess}, \citenamefont {Stahl}, \citenamefont
  {Kenworthy}, \citenamefont {Pluck}, \citenamefont {Macoretta}, \citenamefont
  {Brout}, \citenamefont {Jones}, \citenamefont {Scolnic},\ and\ \citenamefont
  {Filippenko}}]{murakami2023leveraging}%
  \BibitemOpen
  \bibfield  {author} {\bibinfo {author} {\bibfnamefont {Y.~S.}\ \bibnamefont
  {Murakami}}, \bibinfo {author} {\bibfnamefont {A.~G.}\ \bibnamefont {Riess}},
  \bibinfo {author} {\bibfnamefont {B.~E.}\ \bibnamefont {Stahl}}, \bibinfo
  {author} {\bibfnamefont {W.~D.}\ \bibnamefont {Kenworthy}}, \bibinfo {author}
  {\bibfnamefont {D.-M.~A.}\ \bibnamefont {Pluck}}, \bibinfo {author}
  {\bibfnamefont {A.}~\bibnamefont {Macoretta}}, \bibinfo {author}
  {\bibfnamefont {D.}~\bibnamefont {Brout}}, \bibinfo {author} {\bibfnamefont
  {D.~O.}\ \bibnamefont {Jones}}, \bibinfo {author} {\bibfnamefont {D.~M.}\
  \bibnamefont {Scolnic}},  and \bibinfo {author} {\bibfnamefont {A.~V.}\
  \bibnamefont {Filippenko}},\ }\href@noop {} {\enquote {\bibinfo {title}
  {Leveraging sn ia spectroscopic similarity to improve the measurement of
  $h_0$},}\ } (\bibinfo {year} {2023}),\ \Eprint
  {http://arxiv.org/abs/2306.00070} {arXiv:2306.00070 [astro-ph.CO]}
  \BibitemShut {NoStop}%
\bibitem [{\citenamefont {et~al}(2023)}]{qu2023atacama}%
  \BibitemOpen
  \bibfield  {author} {\bibinfo {author} {\bibfnamefont {F.~J.~Q.}\
  \bibnamefont {et~al}},\ }\href@noop {} {\enquote {\bibinfo {title} {The
  atacama cosmology telescope: A measurement of the dr6 cmb lensing power
  spectrum and its implications for structure growth},}\ } (\bibinfo {year}
  {2023}),\ \Eprint {http://arxiv.org/abs/2304.05202} {arXiv:2304.05202
  [astro-ph.CO]} \BibitemShut {NoStop}%
\bibitem [{\citenamefont {Di~Valentino}\ \emph
  {et~al.}(2021{\natexlab{a}})\citenamefont {Di~Valentino} \emph
  {et~al.}}]{DiValentino:2020vvd}%
  \BibitemOpen
  \bibfield  {author} {\bibinfo {author} {\bibfnamefont {E.}~\bibnamefont
  {Di~Valentino}} \emph {et~al.},\ }\href {\doibase
  10.1016/j.astropartphys.2021.102604} {\bibfield  {journal} {\bibinfo
  {journal} {\emph {Astropart. Phys.}}\ }\textbf {\bibinfo {volume} {131}},\
  \bibinfo {pages} {102604} (\bibinfo {year} {2021}{\natexlab{a}})},\ \Eprint
  {http://arxiv.org/abs/2008.11285} {arXiv:2008.11285 [astro-ph.CO]}
  \BibitemShut {NoStop}%
\bibitem [{\citenamefont {Nunes}\ and\ \citenamefont
  {Vagnozzi}(2021)}]{Nunes_2021}%
  \BibitemOpen
  \bibfield  {author} {\bibinfo {author} {\bibfnamefont {R.~C.}\ \bibnamefont
  {Nunes}} and \bibinfo {author} {\bibfnamefont {S.}~\bibnamefont {Vagnozzi}},\
  }\href {\doibase 10.1093/mnras/stab1613} {\bibfield  {journal} {\bibinfo
  {journal} {\emph {Monthly Notices of the Royal Astronomical Society}}\
  }\textbf {\bibinfo {volume} {505}},\ \bibinfo {pages} {5427} (\bibinfo {year}
  {2021})}\BibitemShut {NoStop}%
\bibitem [{\citenamefont {Amon}\ \emph {et~al.}(2022)\citenamefont {Amon} \emph
  {et~al.}}]{DES:2021bvc}%
  \BibitemOpen
  \bibfield  {author} {\bibinfo {author} {\bibfnamefont {A.}~\bibnamefont
  {Amon}} \emph {et~al.} (\bibinfo {collaboration} {DES}),\ }\href {\doibase
  10.1103/PhysRevD.105.023514} {\bibfield  {journal} {\bibinfo  {journal}
  {\emph {Phys. Rev. D}}\ }\textbf {\bibinfo {volume} {105}},\ \bibinfo {pages}
  {023514} (\bibinfo {year} {2022})},\ \Eprint
  {http://arxiv.org/abs/2105.13543} {arXiv:2105.13543 [astro-ph.CO]}
  \BibitemShut {NoStop}%
\bibitem [{\citenamefont {et~al}(2021)}]{Asgari_2021}%
  \BibitemOpen
  \bibfield  {author} {\bibinfo {author} {\bibfnamefont {M.~A.}\ \bibnamefont
  {et~al}},\ }\href {\doibase 10.1051/0004-6361/202039070} {\bibfield
  {journal} {\bibinfo  {journal} {\emph {Astronomy {\&} Astrophysics}}\
  }\textbf {\bibinfo {volume} {645}},\ \bibinfo {pages} {A104} (\bibinfo {year}
  {2021})}\BibitemShut {NoStop}%
\bibitem [{\citenamefont {Perivolaropoulos}\ and\ \citenamefont
  {Skara}(2022)}]{Perivolaropoulos_2022}%
  \BibitemOpen
  \bibfield  {author} {\bibinfo {author} {\bibfnamefont {L.}~\bibnamefont
  {Perivolaropoulos}} and \bibinfo {author} {\bibfnamefont {F.}~\bibnamefont
  {Skara}},\ }\href {\doibase 10.1016/j.newar.2022.101659} {\bibfield
  {journal} {\bibinfo  {journal} {\emph {New Astronomy Reviews}}\ }\textbf
  {\bibinfo {volume} {95}},\ \bibinfo {pages} {101659} (\bibinfo {year}
  {2022})}\BibitemShut {NoStop}%
\bibitem [{\citenamefont {Di~Valentino}\ \emph
  {et~al.}(2021{\natexlab{b}})\citenamefont {Di~Valentino}, \citenamefont
  {Mena}, \citenamefont {Pan}, \citenamefont {Visinelli}, \citenamefont {Yang},
  \citenamefont {Melchiorri}, \citenamefont {Mota}, \citenamefont {Riess},\
  and\ \citenamefont {Silk}}]{DiValentino:2021izs}%
  \BibitemOpen
  \bibfield  {author} {\bibinfo {author} {\bibfnamefont {E.}~\bibnamefont
  {Di~Valentino}}, \bibinfo {author} {\bibfnamefont {O.}~\bibnamefont {Mena}},
  \bibinfo {author} {\bibfnamefont {S.}~\bibnamefont {Pan}}, \bibinfo {author}
  {\bibfnamefont {L.}~\bibnamefont {Visinelli}}, \bibinfo {author}
  {\bibfnamefont {W.}~\bibnamefont {Yang}}, \bibinfo {author} {\bibfnamefont
  {A.}~\bibnamefont {Melchiorri}}, \bibinfo {author} {\bibfnamefont {D.~F.}\
  \bibnamefont {Mota}}, \bibinfo {author} {\bibfnamefont {A.~G.}\ \bibnamefont
  {Riess}},  and \bibinfo {author} {\bibfnamefont {J.}~\bibnamefont {Silk}},\
  }\href {\doibase 10.1088/1361-6382/ac086d} {\bibfield  {journal} {\bibinfo
  {journal} {\emph {Class. Quant. Grav.}}\ }\textbf {\bibinfo {volume} {38}},\
  \bibinfo {pages} {153001} (\bibinfo {year} {2021}{\natexlab{b}})},\ \Eprint
  {http://arxiv.org/abs/2103.01183} {arXiv:2103.01183 [astro-ph.CO]}
  \BibitemShut {NoStop}%
\bibitem [{\citenamefont {de~Rham}(2014)}]{dRGT_2}%
  \BibitemOpen
  \bibfield  {author} {\bibinfo {author} {\bibfnamefont {C.}~\bibnamefont
  {de~Rham}},\ }\href {\doibase 10.12942/lrr-2014-7} {\bibfield  {journal}
  {\bibinfo  {journal} {\emph {Living Reviews in Relativity}}\ }\textbf
  {\bibinfo {volume} {17}} (\bibinfo {year} {2014}),\
  10.12942/lrr-2014-7}\BibitemShut {NoStop}%
\bibitem [{\citenamefont {De~Felice}\ and\ \citenamefont
  {Mukohyama}(2016{\natexlab{a}})}]{MTMG:origpap}%
  \BibitemOpen
  \bibfield  {author} {\bibinfo {author} {\bibfnamefont {A.}~\bibnamefont
  {De~Felice}} and \bibinfo {author} {\bibfnamefont {S.}~\bibnamefont
  {Mukohyama}},\ }\href@noop {} {\bibfield  {journal} {\bibinfo  {journal}
  {\emph {Physics Letters B}}\ }\textbf {\bibinfo {volume} {752}},\ \bibinfo
  {pages} {302} (\bibinfo {year} {2016}{\natexlab{a}})},\ \Eprint
  {http://arxiv.org/abs/1506.01594} {arXiv:1506.01594 [gr-qc]} \BibitemShut
  {NoStop}%
\bibitem [{\citenamefont {De~Felice}\ and\ \citenamefont
  {Mukohyama}(2016{\natexlab{b}})}]{MTMG:pheno}%
  \BibitemOpen
  \bibfield  {author} {\bibinfo {author} {\bibfnamefont {A.}~\bibnamefont
  {De~Felice}} and \bibinfo {author} {\bibfnamefont {S.}~\bibnamefont
  {Mukohyama}},\ }\href@noop {} {\bibfield  {journal} {\bibinfo  {journal}
  {\emph {JCAP}}\ }\textbf {\bibinfo {volume} {04}},\ \bibinfo {pages} {028}
  (\bibinfo {year} {2016}{\natexlab{b}})},\ \Eprint
  {http://arxiv.org/abs/1512.04008} {arXiv:1512.04008 [gr-qc]} \BibitemShut
  {NoStop}%
\bibitem [{\citenamefont {de~Araujo}\ \emph {et~al.}(2021)\citenamefont
  {de~Araujo}, \citenamefont {De~Felice}, \citenamefont {Kumar},\ and\
  \citenamefont {Nunes}}]{deAraujo:2021cnd}%
  \BibitemOpen
  \bibfield  {author} {\bibinfo {author} {\bibfnamefont {J.~C.~N.}\
  \bibnamefont {de~Araujo}}, \bibinfo {author} {\bibfnamefont {A.}~\bibnamefont
  {De~Felice}}, \bibinfo {author} {\bibfnamefont {S.}~\bibnamefont {Kumar}},
  and \bibinfo {author} {\bibfnamefont {R.~C.}\ \bibnamefont {Nunes}},\ }\href
  {\doibase 10.1103/PhysRevD.104.104057} {\bibfield  {journal} {\bibinfo
  {journal} {\emph {Phys. Rev. D}}\ }\textbf {\bibinfo {volume} {104}},\
  \bibinfo {pages} {104057} (\bibinfo {year} {2021})},\ \Eprint
  {http://arxiv.org/abs/2106.09595} {arXiv:2106.09595 [astro-ph.CO]}
  \BibitemShut {NoStop}%
\bibitem [{\citenamefont {De~Felice}\ \emph
  {et~al.}(2021{\natexlab{a}})\citenamefont {De~Felice}, \citenamefont
  {Mukohyama},\ and\ \citenamefont {Pookkillath}}]{DeFelice:2021trp}%
  \BibitemOpen
  \bibfield  {author} {\bibinfo {author} {\bibfnamefont {A.}~\bibnamefont
  {De~Felice}}, \bibinfo {author} {\bibfnamefont {S.}~\bibnamefont
  {Mukohyama}},  and \bibinfo {author} {\bibfnamefont {M.~C.}\ \bibnamefont
  {Pookkillath}},\ }\href {\doibase 10.1088/1475-7516/2021/12/011} {\bibfield
  {journal} {\bibinfo  {journal} {\emph {JCAP}}\ }\textbf {\bibinfo {volume}
  {12}},\ \bibinfo {pages} {011} (\bibinfo {year} {2021}{\natexlab{a}})},\
  \Eprint {http://arxiv.org/abs/2110.01237} {arXiv:2110.01237 [astro-ph.CO]}
  \BibitemShut {NoStop}%
\bibitem [{\citenamefont {Bolis}\ \emph {et~al.}(2018)\citenamefont {Bolis},
  \citenamefont {De~Felice},\ and\ \citenamefont {Mukohyama}}]{MTMG:isw}%
  \BibitemOpen
  \bibfield  {author} {\bibinfo {author} {\bibfnamefont {N.}~\bibnamefont
  {Bolis}}, \bibinfo {author} {\bibfnamefont {A.}~\bibnamefont {De~Felice}},
  and \bibinfo {author} {\bibfnamefont {S.}~\bibnamefont {Mukohyama}},\ }\href
  {\doibase 10.1103/physrevd.98.024010} {\bibfield  {journal} {\bibinfo
  {journal} {\emph {Phys. Rev. D}}\ }\textbf {\bibinfo {volume} {98}} (\bibinfo
  {year} {2018}),\ 10.1103/physrevd.98.024010},\ \Eprint
  {http://arxiv.org/abs/1804.01790} {arXiv:1804.01790 [astro-ph]} \BibitemShut
  {NoStop}%
\bibitem [{\citenamefont {De~Felice}\ \emph {et~al.}(2018)\citenamefont
  {De~Felice}, \citenamefont {Larrouturou}, \citenamefont {Mukohyama},\ and\
  \citenamefont {Oliosi}}]{MTMG:BH}%
  \BibitemOpen
  \bibfield  {author} {\bibinfo {author} {\bibfnamefont {A.}~\bibnamefont
  {De~Felice}}, \bibinfo {author} {\bibfnamefont {F.}~\bibnamefont
  {Larrouturou}}, \bibinfo {author} {\bibfnamefont {S.}~\bibnamefont
  {Mukohyama}},  and \bibinfo {author} {\bibfnamefont {M.}~\bibnamefont
  {Oliosi}},\ }\href@noop {} {\bibfield  {journal} {\bibinfo  {journal} {\emph
  {Phys. Rev. D}}\ }\textbf {\bibinfo {volume} {98}},\ \bibinfo {pages}
  {104031} (\bibinfo {year} {2018})},\ \Eprint
  {http://arxiv.org/abs/1808.01403} {arXiv:1808.01403 [gr-qc]} \BibitemShut
  {NoStop}%
\bibitem [{\citenamefont {De~Felice}\ \emph
  {et~al.}(2017{\natexlab{a}})\citenamefont {De~Felice}, \citenamefont
  {Mukohyama},\ and\ \citenamefont {Oliosi}}]{MQD_OP}%
  \BibitemOpen
  \bibfield  {author} {\bibinfo {author} {\bibfnamefont {A.}~\bibnamefont
  {De~Felice}}, \bibinfo {author} {\bibfnamefont {S.}~\bibnamefont
  {Mukohyama}},  and \bibinfo {author} {\bibfnamefont {M.}~\bibnamefont
  {Oliosi}},\ }\href {\doibase 10.1103/physrevd.96.024032} {\bibfield
  {journal} {\bibinfo  {journal} {\emph {Phys. Rev. D}}\ }\textbf {\bibinfo
  {volume} {96}} (\bibinfo {year} {2017}{\natexlab{a}}),\
  10.1103/physrevd.96.024032},\ \Eprint {http://arxiv.org/abs/1701.01581}
  {arXiv:1701.01581 [hep-th]} \BibitemShut {NoStop}%
\bibitem [{\citenamefont {De~Felice}\ \emph
  {et~al.}(2017{\natexlab{b}})\citenamefont {De~Felice}, \citenamefont
  {Mukohyama},\ and\ \citenamefont {Oliosi}}]{MQD_Horndeski}%
  \BibitemOpen
  \bibfield  {author} {\bibinfo {author} {\bibfnamefont {A.}~\bibnamefont
  {De~Felice}}, \bibinfo {author} {\bibfnamefont {S.}~\bibnamefont
  {Mukohyama}},  and \bibinfo {author} {\bibfnamefont {M.}~\bibnamefont
  {Oliosi}},\ }\href {\doibase 10.1103/physrevd.96.104036} {\bibfield
  {journal} {\bibinfo  {journal} {\emph {Phys. Rev. D}}\ }\textbf {\bibinfo
  {volume} {96}} (\bibinfo {year} {2017}{\natexlab{b}}),\
  10.1103/physrevd.96.104036},\ \Eprint {http://arxiv.org/abs/1709.03108}
  {arXiv:1709.03108 [hep-th]} \BibitemShut {NoStop}%
\bibitem [{\citenamefont {De~Felice}\ \emph {et~al.}(2019)\citenamefont
  {De~Felice}, \citenamefont {Mukohyama},\ and\ \citenamefont
  {Oliosi}}]{MQD_pheno}%
  \BibitemOpen
  \bibfield  {author} {\bibinfo {author} {\bibfnamefont {A.}~\bibnamefont
  {De~Felice}}, \bibinfo {author} {\bibfnamefont {S.}~\bibnamefont
  {Mukohyama}},  and \bibinfo {author} {\bibfnamefont {M.}~\bibnamefont
  {Oliosi}},\ }\href {\doibase 10.1103/physrevd.99.044055} {\bibfield
  {journal} {\bibinfo  {journal} {\emph {Physical Review D}}\ }\textbf
  {\bibinfo {volume} {99}} (\bibinfo {year} {2019}),\
  10.1103/physrevd.99.044055},\ \Eprint {http://arxiv.org/abs/1806.00602}
  {arXiv:1806.00602 [hep-th]} \BibitemShut {NoStop}%
\bibitem [{\citenamefont {De~Felice}\ \emph
  {et~al.}(2021{\natexlab{b}})\citenamefont {De~Felice}, \citenamefont
  {Larrouturou}, \citenamefont {Mukohyama},\ and\ \citenamefont
  {Oliosi}}]{DeFelice:2020ecp}%
  \BibitemOpen
  \bibfield  {author} {\bibinfo {author} {\bibfnamefont {A.}~\bibnamefont
  {De~Felice}}, \bibinfo {author} {\bibfnamefont {F.}~\bibnamefont
  {Larrouturou}}, \bibinfo {author} {\bibfnamefont {S.}~\bibnamefont
  {Mukohyama}},  and \bibinfo {author} {\bibfnamefont {M.}~\bibnamefont
  {Oliosi}},\ }\href {\doibase 10.1088/1475-7516/2021/04/015} {\bibfield
  {journal} {\bibinfo  {journal} {\emph {JCAP}}\ }\textbf {\bibinfo {volume}
  {04}},\ \bibinfo {pages} {015} (\bibinfo {year} {2021}{\natexlab{b}})},\
  \Eprint {http://arxiv.org/abs/2012.01073} {arXiv:2012.01073 [gr-qc]}
  \BibitemShut {NoStop}%
\bibitem [{\citenamefont {Lin}\ and\ \citenamefont
  {Mukohyama}(2017)}]{MMTG_OP}%
  \BibitemOpen
  \bibfield  {author} {\bibinfo {author} {\bibfnamefont {C.}~\bibnamefont
  {Lin}} and \bibinfo {author} {\bibfnamefont {S.}~\bibnamefont {Mukohyama}},\
  }\href {\doibase 10.1088/1475-7516/2017/10/033} {\bibfield  {journal}
  {\bibinfo  {journal} {\emph {Journal of Cosmology and Astroparticle
  Physics}}\ }\textbf {\bibinfo {volume} {2017}},\ \bibinfo {pages} {033}
  (\bibinfo {year} {2017})},\ \Eprint {http://arxiv.org/abs/1708.03757}
  {arXiv:1708.03757 [gr-qc]} \BibitemShut {NoStop}%
\bibitem [{\citenamefont {Carballo-Rubio}\ \emph {et~al.}(2018)\citenamefont
  {Carballo-Rubio}, \citenamefont {Filippo},\ and\ \citenamefont
  {Liberati}}]{MMTG_Carballo_Rubio}%
  \BibitemOpen
  \bibfield  {author} {\bibinfo {author} {\bibfnamefont {R.}~\bibnamefont
  {Carballo-Rubio}}, \bibinfo {author} {\bibfnamefont {F.~D.}\ \bibnamefont
  {Filippo}},  and \bibinfo {author} {\bibfnamefont {S.}~\bibnamefont
  {Liberati}},\ }\href {\doibase 10.1088/1475-7516/2018/06/026} {\bibfield
  {journal} {\bibinfo  {journal} {\emph {Journal of Cosmology and Astroparticle
  Physics}}\ }\textbf {\bibinfo {volume} {2018}},\ \bibinfo {pages} {026}
  (\bibinfo {year} {2018})},\ \Eprint {http://arxiv.org/abs/1802.02537}
  {arXiv:1802.02537 [gr-qc]} \BibitemShut {NoStop}%
\bibitem [{\citenamefont {Mukohyama}\ and\ \citenamefont
  {Noui}(2019)}]{MMTG_Hamiltonian}%
  \BibitemOpen
  \bibfield  {author} {\bibinfo {author} {\bibfnamefont {S.}~\bibnamefont
  {Mukohyama}} and \bibinfo {author} {\bibfnamefont {K.}~\bibnamefont {Noui}},\
  }\href {\doibase 10.1088/1475-7516/2019/07/049} {\bibfield  {journal}
  {\bibinfo  {journal} {\emph {Journal of Cosmology and Astroparticle
  Physics}}\ }\textbf {\bibinfo {volume} {2019}},\ \bibinfo {pages} {049}
  (\bibinfo {year} {2019})},\ \Eprint {http://arxiv.org/abs/1905.02000}
  {arXiv:1905.02000 [gr-qc]} \BibitemShut {NoStop}%
\bibitem [{\citenamefont {Aoki}\ \emph {et~al.}(2020)\citenamefont {Aoki},
  \citenamefont {De~Felice}, \citenamefont {Mukohyama}, \citenamefont {Noui},
  \citenamefont {Oliosi},\ and\ \citenamefont {C.~Pookkillath}}]{MMTG_Planck}%
  \BibitemOpen
  \bibfield  {author} {\bibinfo {author} {\bibfnamefont {K.}~\bibnamefont
  {Aoki}}, \bibinfo {author} {\bibfnamefont {A.}~\bibnamefont {De~Felice}},
  \bibinfo {author} {\bibfnamefont {S.}~\bibnamefont {Mukohyama}}, \bibinfo
  {author} {\bibfnamefont {K.}~\bibnamefont {Noui}}, \bibinfo {author}
  {\bibfnamefont {M.}~\bibnamefont {Oliosi}},  and \bibinfo {author}
  {\bibfnamefont {M.}~\bibnamefont {C.~Pookkillath}},\ }\href {\doibase
  10.1140/epjc/s10052-020-8291-1} {\bibfield  {journal} {\bibinfo  {journal}
  {\emph {The European Physical Journal C}}\ }\textbf {\bibinfo {volume} {80}}
  (\bibinfo {year} {2020}),\ 10.1140/epjc/s10052-020-8291-1},\ \Eprint
  {http://arxiv.org/abs/2005.13972} {arXiv:2005.13972 [astro-ph.CO]}
  \BibitemShut {NoStop}%
\bibitem [{\citenamefont {De~Felice}\ \emph
  {et~al.}(2021{\natexlab{c}})\citenamefont {De~Felice}, \citenamefont
  {Mukohyama},\ and\ \citenamefont {Pookkillath}}]{VCDM:solvingH0}%
  \BibitemOpen
  \bibfield  {author} {\bibinfo {author} {\bibfnamefont {A.}~\bibnamefont
  {De~Felice}}, \bibinfo {author} {\bibfnamefont {S.}~\bibnamefont
  {Mukohyama}},  and \bibinfo {author} {\bibfnamefont {M.~C.}\ \bibnamefont
  {Pookkillath}},\ }\href {\doibase 10.1016/j.physletb.2021.136201} {\bibfield
  {journal} {\bibinfo  {journal} {\emph {Phys. Lett. B}}\ }\textbf {\bibinfo
  {volume} {816}},\ \bibinfo {pages} {136201} (\bibinfo {year}
  {2021}{\natexlab{c}})},\ \Eprint {http://arxiv.org/abs/2009.08718}
  {arXiv:2009.08718 [astro-ph.CO]} \BibitemShut {NoStop}%
\bibitem [{\citenamefont {Yao}\ \emph {et~al.}(2020)\citenamefont {Yao},
  \citenamefont {Oliosi}, \citenamefont {Gao},\ and\ \citenamefont
  {Mukohyama}}]{Yao:2020tur}%
  \BibitemOpen
  \bibfield  {author} {\bibinfo {author} {\bibfnamefont {Z.-B.}\ \bibnamefont
  {Yao}}, \bibinfo {author} {\bibfnamefont {M.}~\bibnamefont {Oliosi}},
  \bibinfo {author} {\bibfnamefont {X.}~\bibnamefont {Gao}},  and \bibinfo
  {author} {\bibfnamefont {S.}~\bibnamefont {Mukohyama}},\ }\href@noop {}
  {\enquote {\bibinfo {title} {{Minimally modified gravity with an auxiliary
  constraint: a Hamiltonian construction}},}\ } (\bibinfo {year} {2020}),\
  \Eprint {http://arxiv.org/abs/2011.00805} {arXiv:2011.00805 [gr-qc]}
  \BibitemShut {NoStop}%
\bibitem [{\citenamefont {De~Felice}\ \emph {et~al.}(2022)\citenamefont
  {De~Felice}, \citenamefont {Mukohyama},\ and\ \citenamefont
  {Pookkillath}}]{DeFelice:2022mcd}%
  \BibitemOpen
  \bibfield  {author} {\bibinfo {author} {\bibfnamefont {A.}~\bibnamefont
  {De~Felice}}, \bibinfo {author} {\bibfnamefont {S.}~\bibnamefont
  {Mukohyama}},  and \bibinfo {author} {\bibfnamefont {M.~C.}\ \bibnamefont
  {Pookkillath}},\ }\href {\doibase 10.1103/PhysRevD.106.084050} {\bibfield
  {journal} {\bibinfo  {journal} {\emph {Phys. Rev. D}}\ }\textbf {\bibinfo
  {volume} {106}},\ \bibinfo {pages} {084050} (\bibinfo {year} {2022})},\
  \Eprint {http://arxiv.org/abs/2206.03338} {arXiv:2206.03338 [gr-qc]}
  \BibitemShut {NoStop}%
\bibitem [{\citenamefont {Özgür Akarsu}\ \emph {et~al.}(2021)\citenamefont
  {Özgür Akarsu}, \citenamefont {Kumar}, \citenamefont {Özülker},\ and\
  \citenamefont {Vazquez}}]{Akarsu_2021}%
  \BibitemOpen
  \bibfield  {author} {\bibinfo {author} {\bibnamefont {Özgür Akarsu}},
  \bibinfo {author} {\bibfnamefont {S.}~\bibnamefont {Kumar}}, \bibinfo
  {author} {\bibfnamefont {E.}~\bibnamefont {Özülker}},  and \bibinfo
  {author} {\bibfnamefont {J.~A.}\ \bibnamefont {Vazquez}},\ }\href {\doibase
  10.1103/physrevd.104.123512} {\bibfield  {journal} {\bibinfo  {journal}
  {\emph {Physical Review D}}\ }\textbf {\bibinfo {volume} {104}} (\bibinfo
  {year} {2021}),\ 10.1103/physrevd.104.123512}\BibitemShut {NoStop}%
\bibitem [{\citenamefont {Özgür Akarsu}\ \emph {et~al.}(2023)\citenamefont
  {Özgür Akarsu}, \citenamefont {Kumar}, \citenamefont {Özülker},
  \citenamefont {Vazquez},\ and\ \citenamefont {Yadav}}]{Akarsu_2023}%
  \BibitemOpen
  \bibfield  {author} {\bibinfo {author} {\bibnamefont {Özgür Akarsu}},
  \bibinfo {author} {\bibfnamefont {S.}~\bibnamefont {Kumar}}, \bibinfo
  {author} {\bibfnamefont {E.}~\bibnamefont {Özülker}}, \bibinfo {author}
  {\bibfnamefont {J.~A.}\ \bibnamefont {Vazquez}},  and \bibinfo {author}
  {\bibfnamefont {A.}~\bibnamefont {Yadav}},\ }\href {\doibase
  10.1103/physrevd.108.023513} {\bibfield  {journal} {\bibinfo  {journal}
  {\emph {Physical Review D}}\ }\textbf {\bibinfo {volume} {108}} (\bibinfo
  {year} {2023}),\ 10.1103/physrevd.108.023513}\BibitemShut {NoStop}%
\bibitem [{\citenamefont {Akarsu}\ \emph {et~al.}(2023)\citenamefont {Akarsu},
  \citenamefont {Valentino}, \citenamefont {Kumar}, \citenamefont {Nunes},
  \citenamefont {Vazquez},\ and\ \citenamefont {Yadav}}]{akarsu2023lambdarm}%
  \BibitemOpen
  \bibfield  {author} {\bibinfo {author} {\bibfnamefont {O.}~\bibnamefont
  {Akarsu}}, \bibinfo {author} {\bibfnamefont {E.~D.}\ \bibnamefont
  {Valentino}}, \bibinfo {author} {\bibfnamefont {S.}~\bibnamefont {Kumar}},
  \bibinfo {author} {\bibfnamefont {R.~C.}\ \bibnamefont {Nunes}}, \bibinfo
  {author} {\bibfnamefont {J.~A.}\ \bibnamefont {Vazquez}},  and \bibinfo
  {author} {\bibfnamefont {A.}~\bibnamefont {Yadav}},\ }\href@noop {} {\enquote
  {\bibinfo {title} {$\lambda_{\rm s}$cdm model: A promising scenario for
  alleviation of cosmological tensions},}\ } (\bibinfo {year} {2023}),\ \Eprint
  {http://arxiv.org/abs/2307.10899} {arXiv:2307.10899 [astro-ph.CO]}
  \BibitemShut {NoStop}%
\bibitem [{\citenamefont {de~Rham}\ \emph {et~al.}(2011)\citenamefont
  {de~Rham}, \citenamefont {Gabadadze},\ and\ \citenamefont {Tolley}}]{dRGT_1}%
  \BibitemOpen
  \bibfield  {author} {\bibinfo {author} {\bibfnamefont {C.}~\bibnamefont
  {de~Rham}}, \bibinfo {author} {\bibfnamefont {G.}~\bibnamefont {Gabadadze}},
  and \bibinfo {author} {\bibfnamefont {A.~J.}\ \bibnamefont {Tolley}},\ }\href
  {\doibase 10.1103/physrevlett.106.231101} {\bibfield  {journal} {\bibinfo
  {journal} {\emph {Physical Review Letters}}\ }\textbf {\bibinfo {volume}
  {106}} (\bibinfo {year} {2011}),\ 10.1103/physrevlett.106.231101},\ \Eprint
  {http://arxiv.org/abs/1011.1232} {arXiv:1011.1232 [hep-th]} \BibitemShut
  {NoStop}%
\bibitem [{\citenamefont {De~Felice}\ \emph {et~al.}(2012)\citenamefont
  {De~Felice}, \citenamefont {Gumrukcuoglu},\ and\ \citenamefont
  {Mukohyama}}]{DeFelice:2012mx}%
  \BibitemOpen
  \bibfield  {author} {\bibinfo {author} {\bibfnamefont {A.}~\bibnamefont
  {De~Felice}}, \bibinfo {author} {\bibfnamefont {A.~E.}\ \bibnamefont
  {Gumrukcuoglu}},  and \bibinfo {author} {\bibfnamefont {S.}~\bibnamefont
  {Mukohyama}},\ }\href {\doibase 10.1103/PhysRevLett.109.171101} {\bibfield
  {journal} {\bibinfo  {journal} {\emph {Phys. Rev. Lett.}}\ }\textbf {\bibinfo
  {volume} {109}},\ \bibinfo {pages} {171101} (\bibinfo {year} {2012})},\
  \Eprint {http://arxiv.org/abs/1206.2080} {arXiv:1206.2080 [hep-th]}
  \BibitemShut {NoStop}%
\bibitem [{\citenamefont {De~Felice}\ \emph {et~al.}(2013)\citenamefont
  {De~Felice}, \citenamefont {G\"umr\"uk\c{c}\"uo\u{g}lu}, \citenamefont
  {Lin},\ and\ \citenamefont {Mukohyama}}]{DeFelice:2013awa}%
  \BibitemOpen
  \bibfield  {author} {\bibinfo {author} {\bibfnamefont {A.}~\bibnamefont
  {De~Felice}}, \bibinfo {author} {\bibfnamefont {A.~E.}\ \bibnamefont
  {G\"umr\"uk\c{c}\"uo\u{g}lu}}, \bibinfo {author} {\bibfnamefont
  {C.}~\bibnamefont {Lin}},  and \bibinfo {author} {\bibfnamefont
  {S.}~\bibnamefont {Mukohyama}},\ }\href {\doibase
  10.1088/1475-7516/2013/05/035} {\bibfield  {journal} {\bibinfo  {journal}
  {\emph {JCAP}}\ }\textbf {\bibinfo {volume} {05}},\ \bibinfo {pages} {035}
  (\bibinfo {year} {2013})},\ \Eprint {http://arxiv.org/abs/1303.4154}
  {arXiv:1303.4154 [hep-th]} \BibitemShut {NoStop}%
\bibitem [{\citenamefont {Schutz}\ and\ \citenamefont
  {Sorkin}(1977)}]{Schutz:1977df}%
  \BibitemOpen
  \bibfield  {author} {\bibinfo {author} {\bibfnamefont {B.~F.}\ \bibnamefont
  {Schutz}} and \bibinfo {author} {\bibfnamefont {R.}~\bibnamefont {Sorkin}},\
  }\href {\doibase 10.1016/0003-4916(77)90200-7} {\bibfield  {journal}
  {\bibinfo  {journal} {\emph {Annals Phys.}}\ }\textbf {\bibinfo {volume}
  {107}},\ \bibinfo {pages} {1} (\bibinfo {year} {1977})}\BibitemShut {NoStop}%
\bibitem [{\citenamefont {De~Felice}\ and\ \citenamefont
  {Mukohyama}(2016{\natexlab{c}})}]{DeFelice:2015moy}%
  \BibitemOpen
  \bibfield  {author} {\bibinfo {author} {\bibfnamefont {A.}~\bibnamefont
  {De~Felice}} and \bibinfo {author} {\bibfnamefont {S.}~\bibnamefont
  {Mukohyama}},\ }\href {\doibase 10.1088/1475-7516/2016/04/028} {\bibfield
  {journal} {\bibinfo  {journal} {\emph {JCAP}}\ }\textbf {\bibinfo {volume}
  {04}},\ \bibinfo {pages} {028} (\bibinfo {year} {2016}{\natexlab{c}})},\
  \Eprint {http://arxiv.org/abs/1512.04008} {arXiv:1512.04008 [hep-th]}
  \BibitemShut {NoStop}%
\bibitem [{\citenamefont {De~Felice}\ \emph
  {et~al.}(2021{\natexlab{d}})\citenamefont {De~Felice}, \citenamefont
  {Mukohyama},\ and\ \citenamefont {Pookkillath}}]{DeFelice:2020cpt}%
  \BibitemOpen
  \bibfield  {author} {\bibinfo {author} {\bibfnamefont {A.}~\bibnamefont
  {De~Felice}}, \bibinfo {author} {\bibfnamefont {S.}~\bibnamefont
  {Mukohyama}},  and \bibinfo {author} {\bibfnamefont {M.~C.}\ \bibnamefont
  {Pookkillath}},\ }\href {\doibase 10.1016/j.physletb.2021.136201} {\bibfield
  {journal} {\bibinfo  {journal} {\emph {Phys. Lett. B}}\ }\textbf {\bibinfo
  {volume} {816}},\ \bibinfo {pages} {136201} (\bibinfo {year}
  {2021}{\natexlab{d}})},\ \Eprint {http://arxiv.org/abs/2009.08718}
  {arXiv:2009.08718 [astro-ph.CO]} \BibitemShut {NoStop}%
\bibitem [{\citenamefont
  {Collaboration}(2020{\natexlab{a}})}]{Planck:2019nip_a}%
  \BibitemOpen
  \bibfield  {author} {\bibinfo {author} {\bibfnamefont {P.}~\bibnamefont
  {Collaboration}},\ }\href {\doibase 10.1051/0004-6361/201936386} {\bibfield
  {journal} {\bibinfo  {journal} {\emph {Astronomy {\&} Astrophysics}}\
  }\textbf {\bibinfo {volume} {641}},\ \bibinfo {pages} {A5} (\bibinfo {year}
  {2020}{\natexlab{a}})}\BibitemShut {NoStop}%
\bibitem [{\citenamefont
  {Collaboration}(2020{\natexlab{b}})}]{Planck:2018lbu_b}%
  \BibitemOpen
  \bibfield  {author} {\bibinfo {author} {\bibfnamefont {P.}~\bibnamefont
  {Collaboration}},\ }\href {\doibase 10.1051/0004-6361/201833886} {\bibfield
  {journal} {\bibinfo  {journal} {\emph {Astronomy {\&} Astrophysics}}\
  }\textbf {\bibinfo {volume} {641}},\ \bibinfo {pages} {A8} (\bibinfo {year}
  {2020}{\natexlab{b}})}\BibitemShut {NoStop}%
\bibitem [{\citenamefont {Alam}\ \emph {et~al.}(2020)\citenamefont {Alam} \emph
  {et~al.}}]{Alam:2020sor}%
  \BibitemOpen
  \bibfield  {author} {\bibinfo {author} {\bibfnamefont {S.}~\bibnamefont
  {Alam}} \emph {et~al.} (\bibinfo {collaboration} {eBOSS}),\ }\Eprint
  {http://arxiv.org/abs/2007.08991} {arXiv:2007.08991 [astro-ph.CO]}
  \BibitemShut {NoStop}%
\bibitem [{\citenamefont {Brout}\ \emph {et~al.}(2022)\citenamefont {Brout}
  \emph {et~al.}}]{Brout:2022vxf}%
  \BibitemOpen
  \bibfield  {author} {\bibinfo {author} {\bibfnamefont {D.}~\bibnamefont
  {Brout}} \emph {et~al.},\ }\href {\doibase 10.3847/1538-4357/ac8e04}
  {\bibfield  {journal} {\bibinfo  {journal} {\emph {Astrophys. J.}}\ }\textbf
  {\bibinfo {volume} {938}},\ \bibinfo {pages} {110} (\bibinfo {year}
  {2022})},\ \Eprint {http://arxiv.org/abs/2202.04077} {arXiv:2202.04077
  [astro-ph.CO]} \BibitemShut {NoStop}%
\bibitem [{\citenamefont {Kuijken}\ \emph {et~al.}(2019)\citenamefont {Kuijken}
  \emph {et~al.}}]{Kuijken:2019gsa}%
  \BibitemOpen
  \bibfield  {author} {\bibinfo {author} {\bibfnamefont {K.}~\bibnamefont
  {Kuijken}} \emph {et~al.},\ }\href {\doibase 10.1051/0004-6361/201834918}
  {\bibfield  {journal} {\bibinfo  {journal} {\emph {Astron. Astrophys.}}\
  }\textbf {\bibinfo {volume} {625}},\ \bibinfo {pages} {A2} (\bibinfo {year}
  {2019})},\ \Eprint {http://arxiv.org/abs/1902.11265} {arXiv:1902.11265
  [astro-ph.GA]} \BibitemShut {NoStop}%
\bibitem [{\citenamefont {Giblin}\ \emph {et~al.}(2021)\citenamefont {Giblin}
  \emph {et~al.}}]{Giblin:2020quj}%
  \BibitemOpen
  \bibfield  {author} {\bibinfo {author} {\bibfnamefont {B.}~\bibnamefont
  {Giblin}} \emph {et~al.},\ }\href {\doibase 10.1051/0004-6361/202038850}
  {\bibfield  {journal} {\bibinfo  {journal} {\emph {Astron. Astrophys.}}\
  }\textbf {\bibinfo {volume} {645}},\ \bibinfo {pages} {A105} (\bibinfo {year}
  {2021})},\ \Eprint {http://arxiv.org/abs/2007.01845} {arXiv:2007.01845
  [astro-ph.CO]} \BibitemShut {NoStop}%
\bibitem [{\citenamefont {Hildebrandt}\ \emph {et~al.}(2021)\citenamefont
  {Hildebrandt} \emph {et~al.}}]{Hildebrandt:2020rno}%
  \BibitemOpen
  \bibfield  {author} {\bibinfo {author} {\bibfnamefont {H.}~\bibnamefont
  {Hildebrandt}} \emph {et~al.},\ }\href {\doibase 10.1051/0004-6361/202039018}
  {\bibfield  {journal} {\bibinfo  {journal} {\emph {Astron. Astrophys.}}\
  }\textbf {\bibinfo {volume} {647}},\ \bibinfo {pages} {A124} (\bibinfo {year}
  {2021})},\ \Eprint {http://arxiv.org/abs/2007.15635} {arXiv:2007.15635
  [astro-ph.CO]} \BibitemShut {NoStop}%
\bibitem [{\citenamefont {Asgari}\ \emph {et~al.}(2021)\citenamefont {Asgari}
  \emph {et~al.}}]{KiDS:2020suj}%
  \BibitemOpen
  \bibfield  {author} {\bibinfo {author} {\bibfnamefont {M.}~\bibnamefont
  {Asgari}} \emph {et~al.} (\bibinfo {collaboration} {KiDS}),\ }\href {\doibase
  10.1051/0004-6361/202039070} {\bibfield  {journal} {\bibinfo  {journal}
  {\emph {Astron. Astrophys.}}\ }\textbf {\bibinfo {volume} {645}},\ \bibinfo
  {pages} {A104} (\bibinfo {year} {2021})},\ \Eprint
  {http://arxiv.org/abs/2007.15633} {arXiv:2007.15633 [astro-ph.CO]}
  \BibitemShut {NoStop}%
\bibitem [{\citenamefont {Mead}\ \emph {et~al.}(2015)\citenamefont {Mead},
  \citenamefont {Peacock}, \citenamefont {Heymans}, \citenamefont {Joudaki},\
  and\ \citenamefont {Heavens}}]{Mead_2015}%
  \BibitemOpen
  \bibfield  {author} {\bibinfo {author} {\bibfnamefont {A.~J.}\ \bibnamefont
  {Mead}}, \bibinfo {author} {\bibfnamefont {J.~A.}\ \bibnamefont {Peacock}},
  \bibinfo {author} {\bibfnamefont {C.}~\bibnamefont {Heymans}}, \bibinfo
  {author} {\bibfnamefont {S.}~\bibnamefont {Joudaki}},  and \bibinfo {author}
  {\bibfnamefont {A.~F.}\ \bibnamefont {Heavens}},\ }\href {\doibase
  10.1093/mnras/stv2036} {\bibfield  {journal} {\bibinfo  {journal} {\emph
  {Monthly Notices of the Royal Astronomical Society}}\ }\textbf {\bibinfo
  {volume} {454}},\ \bibinfo {pages} {1958} (\bibinfo {year}
  {2015})}\BibitemShut {NoStop}%
\bibitem [{\citenamefont {Hagala}\ \emph {et~al.}(2021)\citenamefont {Hagala},
  \citenamefont {Felice}, \citenamefont {Mota},\ and\ \citenamefont
  {Mukohyama}}]{Hagala:2020eax}%
  \BibitemOpen
  \bibfield  {author} {\bibinfo {author} {\bibfnamefont {R.}~\bibnamefont
  {Hagala}}, \bibinfo {author} {\bibfnamefont {A.~D.}\ \bibnamefont {Felice}},
  \bibinfo {author} {\bibfnamefont {D.~F.}\ \bibnamefont {Mota}},  and \bibinfo
  {author} {\bibfnamefont {S.}~\bibnamefont {Mukohyama}},\ }\href {\doibase
  10.1051/0004-6361/202040018} {\bibfield  {journal} {\bibinfo  {journal}
  {\emph {Astron. Astrophys.}}\ }\textbf {\bibinfo {volume} {653}},\ \bibinfo
  {pages} {A148} (\bibinfo {year} {2021})},\ \Eprint
  {http://arxiv.org/abs/2011.14697} {arXiv:2011.14697 [astro-ph.CO]}
  \BibitemShut {NoStop}%
\bibitem [{\citenamefont {Blas}\ \emph {et~al.}(2011)\citenamefont {Blas},
  \citenamefont {Lesgourgues},\ and\ \citenamefont {Tram}}]{Blas:2011rf}%
  \BibitemOpen
  \bibfield  {author} {\bibinfo {author} {\bibfnamefont {D.}~\bibnamefont
  {Blas}}, \bibinfo {author} {\bibfnamefont {J.}~\bibnamefont {Lesgourgues}},
  and \bibinfo {author} {\bibfnamefont {T.}~\bibnamefont {Tram}},\ }\href
  {\doibase 10.1088/1475-7516/2011/07/034} {\bibfield  {journal} {\bibinfo
  {journal} {\emph {JCAP}}\ }\textbf {\bibinfo {volume} {07}},\ \bibinfo
  {pages} {034} (\bibinfo {year} {2011})},\ \Eprint
  {http://arxiv.org/abs/1104.2933} {arXiv:1104.2933 [astro-ph.CO]} \BibitemShut
  {NoStop}%
\bibitem [{\citenamefont {Audren}\ \emph {et~al.}(2013)\citenamefont {Audren},
  \citenamefont {Lesgourgues}, \citenamefont {Benabed},\ and\ \citenamefont
  {Prunet}}]{Audren:2012wb}%
  \BibitemOpen
  \bibfield  {author} {\bibinfo {author} {\bibfnamefont {B.}~\bibnamefont
  {Audren}}, \bibinfo {author} {\bibfnamefont {J.}~\bibnamefont {Lesgourgues}},
  \bibinfo {author} {\bibfnamefont {K.}~\bibnamefont {Benabed}},  and \bibinfo
  {author} {\bibfnamefont {S.}~\bibnamefont {Prunet}},\ }\href {\doibase
  10.1088/1475-7516/2013/02/001} {\bibfield  {journal} {\bibinfo  {journal}
  {\emph {JCAP}}\ }\textbf {\bibinfo {volume} {02}},\ \bibinfo {pages} {001}
  (\bibinfo {year} {2013})},\ \Eprint {http://arxiv.org/abs/1210.7183}
  {arXiv:1210.7183 [astro-ph.CO]} \BibitemShut {NoStop}%
\bibitem [{\citenamefont {Brinckmann}\ and\ \citenamefont
  {Lesgourgues}(2019)}]{Brinckmann:2018cvx}%
  \BibitemOpen
  \bibfield  {author} {\bibinfo {author} {\bibfnamefont {T.}~\bibnamefont
  {Brinckmann}} and \bibinfo {author} {\bibfnamefont {J.}~\bibnamefont
  {Lesgourgues}},\ }\href {\doibase 10.1016/j.dark.2018.100260} {\bibfield
  {journal} {\bibinfo  {journal} {\emph {Phys. Dark Univ.}}\ }\textbf {\bibinfo
  {volume} {24}},\ \bibinfo {pages} {100260} (\bibinfo {year} {2019})},\
  \Eprint {http://arxiv.org/abs/1804.07261} {arXiv:1804.07261 [astro-ph.CO]}
  \BibitemShut {NoStop}%
\end{thebibliography}%

\end{document}